\RequirePackage{fix-cm}
\documentclass[smallextended]{svjour3}       % onecolumn (second format)
\smartqed  % flush right qed marks, e.g. at end of proof
\usepackage{graphicx}
\usepackage{hyperref}

\usepackage{graphicx}% Include figure files
\usepackage{dcolumn}% Align table columns on decimal point
\usepackage{bm}% bold math

\usepackage{booktabs}
\usepackage{topcapt}

\usepackage{xcolor}

\usepackage{slashed} 
\usepackage{amsmath}
\usepackage{amssymb}
 \usepackage{axodraw2}
 
 \usepackage{ulem}
 
 \newcommand{\propagator}[2]{\frac{\delta^2 W}{\delta #1 ~ \delta #2}}
 \newcommand{\oneparticle}[2]{\frac{\delta^{#1} \Gamma}{ #2}}
 \newcommand{\connect}[2]{\frac{\delta^{#1} W}{ #2}}
\begin{document}

\title{Looking at QED with Dyson-Schwinger equations: basic equations, Ward-Takahashi identities and the two-photon-two-fermion irreducible vertex}% Force line breaks with \\

\author{Orlando Oliveira    \and Helena Lessa Macedo \and Rodrigo Carmo Terin}


\institute{Orlando Oliveira \at
              CFisUC, Departament of Physics, University of Coimbra, 3004-516 Coimbra, Portugal \\
              \email{orlando@uc.pt}
           \and
           Helena Lessa Macedo \at
               CFisUC, Departament of Physics, University of Coimbra, 3004-516 Coimbra, Portugal \\
                \email{hlessamacedo@gmail.com}
           \and 
           Rodrigo Carmo Terin \at
           Instituto Tecnol\'ogico de Aeron\' autica, DCTA, 12.228-900 S\~ao Jos\' e dos Campos, Brazil \\
           \email{rodrigoterin3003@gmail.com}
}

\maketitle

\begin{abstract}
A minimal truncated set of the integral Dyson-Schwinger equations, in Minkowski spacetime, that allows to explore QED beyond its perturbative solution
is derived for general linear covariant gauges. 
The minimal set includes the equations for the fermion and photon propagators, the photon-fermion vertex, and 
the two-photon-two-fermion one-particle-irreducible diagram.
If the first three equations are exact, to build a closed set of equations, the two-photon-two-fermion equation is truncated ignoring the 
contribution of Green functions with large number of external legs.
%The set of equations is compatible with the renormalization program for QED.
 It is shown that the truncated equation for the two-photon-two-fermion 
vertex reproduces the lowest-order perturbative result in the limit of the small coupling constant. Furthermore, this equation allows to define 
an iterative procedure to compute higher order corrections in the coupling constant. 
The Ward-Takahashi identity for the two-photon-two-fermion irreducible vertex is derived and solved in the soft photon limit, where one of the
photon momenta vanish, in the low photon momenta limit and for general kinematics. 
The solution of the Ward-Takahashi identity determines the longitudinal component of the two-photon-two-fermion irreducible vertex, 
while it is proposed to use the Dyson-Schwinger equation to determine the transverse part of this 
irreducible diagram. The two-photon-two-fermion DSE is solved in heavy fermion limit, considering a simplified version of the QED vertices.
The contribution of this irreducible vertex to a low-energy effective photon-fermion vertex is discussed and the fermionic operators that are 
generated are computed in terms of the fermion propagator functions.
\end{abstract}

\maketitle

\tableofcontents

%====================================================================
\section{Introduction and Motivation}

Quantum Electrodynamics (QED) is one of the first quantum field theories encountered by a newcomer to the subject. 
The standard textbook approach to solve QED is  perturbative theory  \cite{Peskin:1995ev}. 
However, QED can be solved with different techniques as 
the Dyson-Schwinger equations (DSE) \cite{Dyson:1949ha,Schwinger:1951ex,Schwinger:1951hq}, 
 that define  an infinite tower of integral equations, 
 its lattice regularized formulation \cite{Wilson:1974sk} via Monte Carlo simulations,
 see e.g. \cite{Loveridge:2021qzs,Loveridge:2021wav,Loveridge:2022sih} and references,  
or using the functional renormalization group approach, see e.g. \cite{Gies:2006wv} for  introduction and references.
In all cases,  perturbation theory  reappears as a special case when the solution is  written
as a power series in the coupling constant. 
Still, there are several features of the dynamics of a quantum field theory (QFT) that cannot be explained 
within a perturbative solution. This is the case, for example, of the dynamical generation of masses in QED
 \cite{Schwinger:1962tn,Polyakov:1976fu,Atkinson:1989fp,Rojas:2008sg,Fischer:2004nq,Kizilersu:2014ela,Loveridge:2021qzs}, 
 that also occurs in QCD \cite{Pagels:1978ba,Oliveira:2018lln},  together with confinement  \cite{Alkofer:2010ue,Loveridge:2022sih}, 
 or the formation of bound states. 
 
The Dyson-Schwinger equations are a set of integral equations relating the full set of the QED Green functions \cite{Roberts:1994dr}. 
From a practical point of view, it is impossible to handle simultaneously all the Green functions and only a subset can be investigated at a time. 
In principle, by including as many as possible Green functions in the analysis of the DSE, one becomes closer and closer to the exact solution for QED
\cite{Bender:2022eze}.
Of the DSE for QED it is the equation for the fermion propagator, also known as the gap equation, 
together with the equation for the photon propagator that have been most investigated in the literature, 
However, to build a solution for these equations the photon-fermion vertex needs to be provided
\cite{Roberts:1994dr,Marciano:1977su,Alkofer:2000wg}.
For example, for a tree level like photon-fermion vertex, the fermionic gap equation allows for the dynamical generation of a fermion mass
if the coupling constant is sufficiently large  \cite{Atkinson:1989fp,Rojas:2008sg,Kizilersu:2014ela}.
The corresponding equation for the photon propagator has a massive solution if the aforementioned vertex is singular at zero momentum 
\cite{Schwinger:1962tn,Cornwall:1981zr,Aguilar:2010cn}.
In a way, the characteristics of the solutions for the propagators are a function of the dressing of the photon-fermion vertex 
used \cite{Curtis:1990zs,Bashir:1997qt}. 
Typically, in QED the photon-fermion vertex is built by solving the vertex Ward-Takahashi identity (WTI) and adding perturbative corrections.
However, the WTI does not provides a complete description of the photon-fermion vertex as it only its longitudinal component relative
to the photon momentum \cite{Guzman:2023hzq}.

The DSE for the photon-fermion vertex requires the two-photon-two-fermion vertex, whose DSE calls for further Green functions with 
larger number of external legs. 
From the point of view of building a solution  of QED that is written as a power series in $\alpha$, these higher-order
Green functions appear as corrections that are associated with higher powers of the coupling constant 
$\alpha$ and, for a small enough coupling constant, they are suppressed.
However, by relying on an enlarged set of DSE, at least a finite set of these would-be corrections 
can be considered in the solution of the theory.
Our main goal is to solve QED via a minimal set of DSE that go beyond the propagator equations.
Eventually, this will allow to improve the kernel that is used to describe the phenomenology associated with QED 
that uses the Bethe-Salpeter or the Faddeev equations to tackle two and three body problems, respectively.

In order to address QED through the Dyson-Schwinger equations and to consider a minimal set of Green functions to solve this QFT, 
we will look only at those Green functions mentioned, i.e. the propagators, the photon-fermion vertex and the two-photon-two-fermion irreducible 
vertex $\Gamma^{\mu \nu}$. No further QED Green functions will be considered in the present work. 
In particular, for the two-photon-two-fermion irreducible vertex its DSE will be derived. 
This integral equation is complex and requires a five point Green functions and, to arrive at a manageable set of equations,
an approximated equation for $\Gamma^{\mu \nu}$ will be built.

The relevance of the two-photon-two-fermion irreducible vertex goes beyond its contribution to the solution of QED. 
Indeed, $\Gamma^{\mu\nu}$
has been considered long ago in the study of non-linear interactions between electromagnetic fields that include light-by-light scattering,
two-quantum pair creation, and the scattering of light by an external electromagnetic field \cite{Karplus:1950zza,Karplus:1950zz}. It is
also responsible for the main contribution of the two-photon exchange corrections to the electron-proton and electron-hadron processes
\cite{Akhmadaliev:1998zz,Arrington:2011dn,{Eichmann:2012mp},Afanasev:2017gsk,Qweak:2021ijt} that determine 
the nucleon and nuclear form factors, it impacts  the electro-production cross section of resonances and pions, 
 the $\gamma Z$ interference in parity-violating electron scattering, etc. 
The two-photon-two-fermion vertex is also relevant in the computation of the muonic-hydrogen Lamb shift  and hyperfine splitting of muonic atoms
\cite{Carlson:2011zd,Antognini:2013rsa,Gorchtein:2014hla,Krauth:2021foz,Fu:2022fgh}, in the calculation of the muon magnetic anomaly, 
see \cite{Keshavarzi:2021eqa} and references therein, and for the extraction of the proton charge radius, see \cite{Gao:2021sml} for a recent review and references therein. 
This irreducible vertex is also relevant to virtual Compton scattering processes \cite{Scherer:1996ux,CLAS:2001wjj,Belitsky:2001ns}
and  the computation and extraction of generalized Parton distributions \cite{Belitsky:2005qn,Shiells:2021xqo,Chavez:2021koz}.
Non-linear electrodynamics effects observed in intense laser beams \cite{Karbstein:2019oej,Blackburn:2019rfv} have also to consider multiple photon signals
and, therefore, must take into account $\Gamma^{\mu\nu}$ and other irreducible diagrams. 
Note that, presently, there are  ongoing experimental programs to
investigate two-photon processes \cite{Rachek:2015ymm,OLYMPUS:2016gso,Abramowicz:2019gvx,CLAS:2021ovm} among other effects.
These experimental programs are an extra motivation to have a closer look at the two-photon-two-fermion irreducible diagrams.

As stated previously, herein, we aim to define a theoretical setup, based on the Dyson-Schwinger equations,
to investigate, from first principles, QED and to go beyond the perturbative approach to Quantum Field Theories looking, in particular, also
to the two-photon-two-fermion vertex. Then, 
to have a consistent notation for the approach that we are aiming for, besides the DSE for the two-photon-two-fermion irreducible vertex, 
the gap equation for the fermion and for the photon propagators, together with the photon-fermion vertex DSE are re-derived. 
For the two-photon-two-fermion Green function, the corresponding DSE is analysed in an approximated (truncated) version that, 
at the lowest order in the coupling constant, reproduces the perturbative result. 
As discussed below, in a perturbative-like solution of this DSE, the truncated integral equation reproduces the lowest order perturbative result, and
it allows the building of an iterative procedure to estimate higher order corrections.
Moreover, we are able to solve, within a particular approximation and in the heavy quark limit, the DSE for the two-photon-two-fermion vertex.

Besides the derivation of the DSE for the Green functions, 
the Ward-Takahashi identities for the two-photon-two-fermion irreducible vertex and for  the photon-fermion vertex are derived and solved. 
For the photon-fermion vertex, the solution of the Ward-Takahashi identity results in the Ball-Chiu photon-fermion vertex \cite{Ball:1980ay}
that takes into account
only longitudinal tensors, relative to the photon momentum, and describes the vertex in terms of the fermion propagator functions.
Similarly as for the photon-fermion vertex, the solution of the WTI for the two-photon-two-fermion irreducible vertex 
that is built  is free of kinematical singularities, it provides only a longitudinal component of this vertex, relative to the photon momenta, 
and writes this component in terms of the fermion propagator functions and of the full photon-fermion vertex.
This, eventually, will allow in the modelling $\Gamma^{\mu\nu}$ in connection with the modelling of the photon-fermion vertex.
Moreover, exact solutions of the WTI identity in the low energy limit are also derived, namely for the case where one of the photon momenta 
is set to zero (soft photon limit) or by taking into account only the linear terms the photon momenta. 

Along the way to achieve the goals described above, i.e., to arrive at the various DSE equations, one needs to investigate the
decomposition of the QED Green functions in terms of connected functions and in terms of one-particle irreducible diagrams. 
The decomposition follows the usual rules of QFT, and its discussion helps to set the notation. In the current manuscript, the reader
can find details on how to perform such decompositions and, using our notation, can find expressions for the various cases that were 
considered.
The motivation to include such information is twofold. To help possible newcomers to the field and to provide details on the decomposition 
of the Green functions that, in some cases, are valid for theories other than QED with minimal modifications.
In the current work,  given the complexity of the underlying mathematical problem, no attempt is made to solve any of the integral equations. 
Indeed, the analysis and computation of the solutions for the DSE is a complex task, that involves various subtleties, and will be the subject of future 
publications. All the computations reported are performed in Minkowski spacetime.

This manuscript is organized as follows. 
In Sec. \ref{Sec:def} we provide the definitions of the various functional generators and some useful relations used throughout the
paper. The Dyson-Schwinger equations for the fermion propagator,
for the photon propagator, and for the photon-fermion vertex are derived in Sec. \ref{Sec:DSE}, while the Ward-Takahashi identities
are computed in Sec. \ref{Sec:WTI}. The solution of the WTI for the photon-fermion vertex, that fixes its longitudinal part, is revisited in 
Sec . \ref{Sec:SolWTI-vertex}. The solution of the Ward-Takahashi identity for the two-photon-two-fermion vertex is discussed in
Sec. \ref{Sec:SolWTI-2P2F}. We give solutions to this WTI in the soft limit, in the linear approximation in the photon momenta (low energy limit)
and for a general kinematics.
The DSE for the two-photon-two-fermion irreducible vertex is derived in Sec. \ref{Sec:VertexfromWTI}, together with its approximate version. 
Furthermore, we show that the approximate DSE  reproduces the results of perturbation theory in the appropriate limit and discuss the implementation 
of  an iterative procedure to improve the lowest-order expression. The problem of building a minimal tensor basis for $\Gamma^{\mu\nu}$ is
touched in Sec. \ref{Sec:2P2F-Tensor-Basis} and an approximate solution to this four-point Green function is built in Sec \ref{SecDSE-2P2F-Sol}.
In Sec. \ref{Sec:EffVert} we discuss the contribution of the two-photon-two-fermion irreducible diagram to an effective low-energy photon-fermion vertex 
and, in particular, what type of effective operators emerge.
The renormalization of QED Dyson-Schwinger equations is discussed in Sec. \ref{Sec:Renor}. 
Finally, Sec. \ref{Sec:Summary} summarizes and concludes. 
In the appendices, we collect several auxiliary results that include the decomposition of the Green functions in terms of one particle irreducible diagrams
and the definitions in momentum space (App.\ref{Sec:Dec}) and the writing of the Ward-Takahashi identities in momentum space (App. \ref{Sec:WTI-Mom}).
Our manuscript includes many computational details that can be skipped in a first reading.

%===========================================================================
%===========================================================================
\section{Quantum Electrodynamics - definitions and generating functionals \label{Sec:def}}

The classical theory of electromagnetism considers the interaction of the four-potential field $A^\mu(x) = ( V(x), \, \vec{A} (x))$, 
where $V$ is the scalar potential and $\vec{A}$ the vector potential, with a Dirac spinorial field $\psi$ that is described by the Lagrangian density
\begin{equation}
\mathcal{L}  =  -\frac{1}{4}F^{\mu\nu}F_{\mu\nu}+\bar{\psi}(x)\Big(i\gamma^{\mu}D_{\mu}-m\Big)\psi(x) \ ,
 \label{1}
\end{equation}
where  the covariant derivative and the Abelian field strength are given by
\begin{eqnarray}
D_{\mu}(x)  =  \partial_{\mu} + i \, g  \, A_{\mu}(x) 
\qquad\mbox{ and }\qquad
   F_{\mu\nu} = \partial_\mu A_\nu - \partial_\nu A_\mu  \ ,
\end{eqnarray}
while $g$ refers to the electric charge.
The  Lagragian density $\mathcal{L}$ is invariant under the gauge transformation
\begin{eqnarray}
& & 
\psi(x)           \rightarrow   e^{i \, \theta(x) } \, \psi(x) \ ,  \qquad 
\bar{\psi}(x)   \rightarrow  \bar{\psi}(x) \, e^{ - \, i \, \theta(x)} 
\nonumber \\
& &
\qquad\mbox{and}\qquad 
A_{\mu}(x)    \rightarrow   A_{\mu}(x) - \frac{1}{g}\partial_{\mu}\theta(x) \ .
\label{Eq:GaugeTransf}
\end{eqnarray}
The set of transformations associated with the fermion field defines the $U(1)$ gauge symmetry group of electromagnetism. 
The definition of the corresponding quantum field theory requires the introduction of a gauge fixing term to build the generating functional for the Green functions $Z[J]$.
The Faddeev-Popov construction results in
\begin{eqnarray}
& &
   Z[J, \bar\eta, \eta]  =  e^{i \, W [J, \bar\eta, \eta] } 
   =   \int \mathcal{D}A \, \mathcal{D}\bar\psi \, \mathcal{D}\psi \, 
   \nonumber \\
   & & 
    \exp\left\{ i  \int d^4 x \Big[ \mathcal{L}_{QED} (A, \bar\psi, \psi) + 
   J^\mu(x)A_\mu(x) + \bar\eta(x) \, \psi(x) + \bar\psi (x)  \, \eta(x)  \Big] \right\}  
   \label{Eq:Geradores}
\end{eqnarray}   
where
\begin{eqnarray}
\mathcal{L}_{QED} & = & \mathcal{L} - \frac{1}{2 \xi} \left( \partial A \right)^2 
\label{Eq:Lqed}
\end{eqnarray}
and the fermion fields $\psi$, $\bar{\psi}$ together with the fermionic sources $\eta$, $\bar\eta$ are  Grassmann fields.
The real parameter $\xi$ is the gauge fixing parameter that defines the linear covariant gauge. The Landau gauge is defined by taking $\xi = 0$.
In addition, $W[J, \bar\eta, \eta]$ is the generating functional of the connected Green functions.

In the functional integration over the fundamental fields $A_\mu$, $\psi$ or $\bar\psi$, these can be replaced by functional 
derivatives\footnote{For the differentiation with respect to the Grassmann numbers we will consider left differentiation.} with respect
to the sources following the rules
\begin{equation}
\psi(x) \leftrightarrow \frac{\delta}{i \, \delta\bar{\eta}(x)} \ ,  \qquad
\bar{\psi}(x) \leftrightarrow \frac{\delta}{-i \, \delta{\eta(x)}} \quad\mbox{and}\quad   A^{\mu}(x)\leftrightarrow\frac{\delta}{i \, \delta J_{\mu}(x)} \ .
\end{equation}
By introducing  the classical fields
\begin{equation}
   A_{c l, \mu}(x) = \frac{\delta W}{\delta J^\mu(x)} \ ,
   \qquad\quad
   \bar\psi_{cl}(x) = - \, \frac{\delta W}{\delta \eta(x)} \ ,
   \qquad\quad
   \psi_{cl}(x) = \frac{\delta W}{\delta \bar\eta(x)} 
   \label{ClFields}
\end{equation}
the generating functional for the one-particle irreducible diagrams is defined via the Legendre transformation
\begin{equation}
  \Gamma [ A_{cl}, \, \bar\psi_{cl} , \, \psi_{cl} ] =
  W[J, \eta, \bar\eta ] - (J,A_{cl}) - (\bar\eta, \psi_{cl}) - (\bar\psi_{cl} , \eta) 
  \label{GammaDef}
\end{equation}  
where
\begin{eqnarray}
& & 
(J,A_{cl})   =  \int d^4 x ~ J^\mu(x) \, A_{cl ,\, \mu}(x) \ , \quad %\\
(\bar\eta, \psi_{cl})  =  \int d^4 x ~ \bar\eta_\alpha (x) \, \psi_{cl , \, \alpha} (x) 
\nonumber \\
& &
\qquad\mbox{and}\qquad %\\
 (\bar\psi_{cl}, \eta)   =  \int d^4 x ~ \bar\psi_{cl , \, \alpha} (x) \, \eta_\alpha (x)  \ .
\end{eqnarray}
It follows from the definitions given in Eqs (\ref{ClFields}) and (\ref{GammaDef}) and the rules of functional derivation that
\begin{equation}
   \frac{\delta\, \Gamma}{\delta A_{cl , \mu} (x)} = - \, J^\mu (x) , 
   \qquad\quad
   \frac{\delta\, \Gamma}{\delta \psi_{cl , \alpha} (x)} =  \bar\eta_\alpha (x) , 
   \qquad\quad
   \frac{\delta \, \Gamma}{\delta \bar\psi_{cl , \alpha} (x)} =  - \, \eta_\alpha (x) .
   \label{DeltaGammas}
\end{equation}
Furthermore, the classical fields are independent fields which are translated into the relations
\begin{eqnarray}
 \int \, d^4 z ~ \frac{\delta^2 W}{\delta J^\mu (x) ~ \delta J^\zeta (z)} ~ \frac{\delta^2 \Gamma}{\delta A_{cl, \, \zeta} (z) ~ \delta A_{cl}^{ \, \nu} (y)} 
                  & = & - ~ g_{\mu\nu} \, \delta ( x -y ) \ ,   \label{OrthoA} \\
 \int \, d^4 z ~ \frac{\delta^2 W}{\delta \eta_\iota (z) ~ \delta \bar\eta_\alpha  (x)} ~ 
                       \frac{\delta^2 \Gamma}{\delta \psi_{cl, \, \beta} (y) ~ \delta \bar\psi_{cl, \, \iota} (z)} 
                &  =  & - ~ \delta_{\alpha\beta} \, \delta ( x -y ) \ ,  \label{Ortho1} \\
 \int \, d^4 z ~ \frac{\delta^2 W}{\delta \bar\eta_\iota (z) ~ \delta \eta_\alpha  (x)} ~ 
                       \frac{\delta^2 \Gamma}{\delta \bar\psi_{cl, \, \beta} (y) ~  \delta \psi_{cl, \, \iota} (z)} 
                &  =  & -  ~\delta_{\alpha\beta} \, \delta ( x -y ) \ ,  \label{Ortho2}
\end{eqnarray}
that hold in the limit where all the sources vanish. The results in Eqs (\ref{OrthoA}) - (\ref{Ortho2}) also connect the second derivatives of
the functional generator $\Gamma$ with the inverse of  the propagators for QED. Indeed, in perturbation theory and at the lowest order in the coupling constant,
\begin{equation}
   W[J, \bar\eta, \eta] = - \, \frac{1}{2} \, \left( J, \, D^{(0)} J \right) - \left( \bar\eta, \, S^{(0)} \eta \right)  + \mathcal{O}(g)  \ ,
   \label{Eq:W-pert}
\end{equation}
where $D^{(0)}$ and $S^{(0)}$ are the tree level photon and fermion propagators, respectively, i.e.
\begin{eqnarray}
 & &
 \hspace{-0.7cm}
\left[ D^{(0)} (x-y) \right]^{\mu\nu} = \int \frac{d^4k}{(2 \, \pi)^4} ~ e^{- \, i \, k \, (x-y) } ~  \frac{1}{k^2 + i \, \epsilon}
\left[ - g^{\mu\nu} + \left( 1 - \xi \right) \frac{k^\mu k^\nu}{k^2}     \right] \ , \\
& &
 \hspace{-0.7cm}
S^{(0)} (x-y) = \int~  \frac{d^4p}{(2 \, \pi)^4}  ~ e^{- \, i \, p \, (x-y) } ~ 
 \frac{ \slashed{p} + m}{p^2 - m^2 + i \, \epsilon} 
\end{eqnarray}
for a fermion of mass $m$. Then, it follows from Eq. (\ref{Eq:W-pert}) that
\begin{eqnarray}
  \left. \frac{\delta^2 W}{\delta J_\mu(x) ~ \delta J_\nu (y) }  \right|_{J=\bar\eta=\eta=0} & = & - \left( D^{(0)} \right)^{ \mu\nu} (x -y )  + \cdots \\
  \left. \frac{\delta^2 W}{\delta \eta_\beta (y) ~ \delta \bar\eta_\alpha (x) }  \right|_{J=\bar\eta=\eta=0} & = & - \, \left. \frac{\delta^2 W}{ \delta \bar\eta_\alpha (x) ~\delta \eta_\beta (y) }  \right|_{J=\bar\eta=\eta=0}  \nonumber \\
   & = & - \left( S^{(0)} \right)_{ \alpha\beta} (x -y ) 
  + \cdots 
\end{eqnarray}
and from Eqs (\ref{OrthoA}) to (\ref{Ortho2}) that the second derivatives of $\Gamma$ are the inverse of the propagators.

Finally and before proceeding further let us comment on the definition of the Landau gauge, that we take as the limit $\xi \rightarrow 0^+$ of the linear covariant
gauges. From the formal point of view, the Landau gauge within this formalism cannot be treated 
since there is no inverse for the photon propagator and, therefore, some
of the formal manipulations are meaningless if one sets $\xi = 0$ from the very beginning. In the following, the expressions for the Landau gauge are therefore derived from those
of the linear covariant gauges after setting the gauge fixing parameter to zero.

%===========================================================================
%===========================================================================
\section{The Dyson-Schwinger Equations \label{Sec:DSE}}

The Dyson-Schwinger equations are the Green functions quantum equations of motion. They can be used to define the theory beyond perturbation theory,
as they are valid for all values of the coupling constants, and are an infinite tower of integral equations relating all QED Green functions. 
In the following, to set the notation and to prepare the discussion of the DSE for the two-photon-two-fermion one-particle irreducible diagram,
the relevant equations will be derived.

%=================================================================
%=================================================================
\subsection{The fermion gap equation}

\begin{figure}[t]
\centering
   \includegraphics[width=3in]{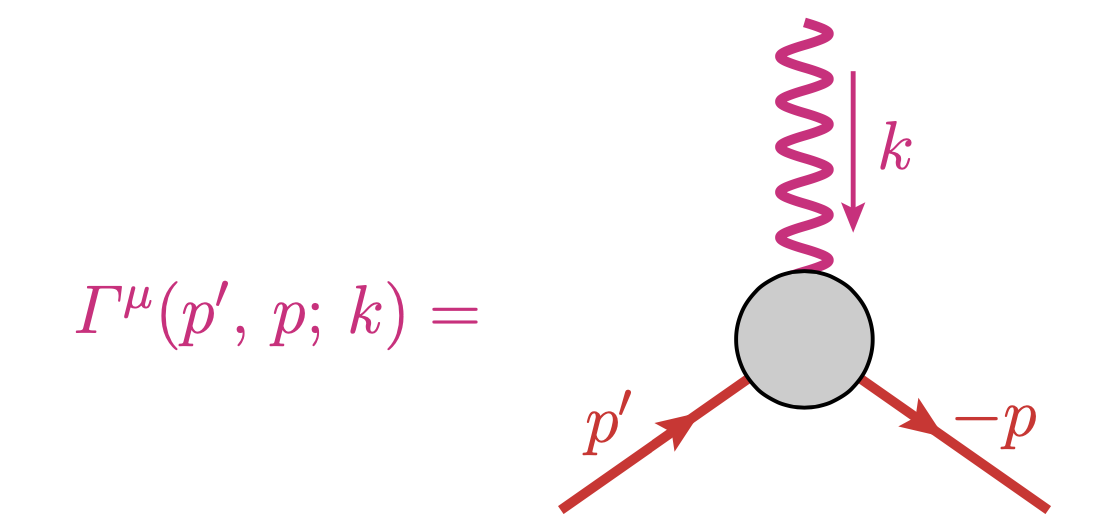} 
\caption{The photon-fermion one-particle irreducible vertex in momentum space: see Eqs  (\ref{DSE-Fermion-realspace}) and (\ref{Eq:Vert_mom}) for definitions. All Feynman diagrams were built using Axodraw \cite{Collins:2016aya}.}
\label{Fig:vertex}
\end{figure}

\begin{figure}[t]
\centering
\includegraphics[width=3.5in]{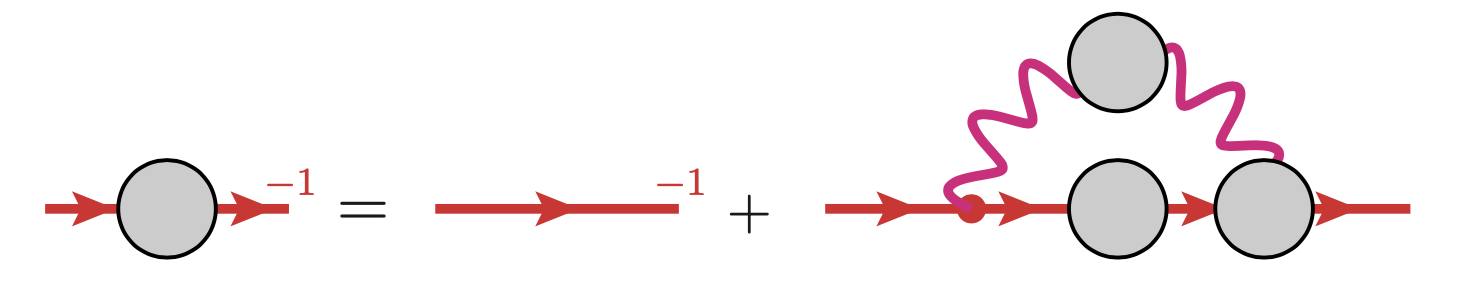} 
\caption{The fermion gap equation. For the vertex the filled blob represents the corresponding one-particule irreducible vertex, while for
the propagators the filled blobs represent full propagators.  The vertex with a dot refers to tree level photon-fermion vertex. Similar notation
is used in all the Figs. in the manuscript.}
\label{Fig:fermiongap}
\end{figure}

The derivation of the DSE relies on the vanishing of the functional integration of the derivatives with respect to the fields. In particular, for the fermion
propagator one starts from the identity
\begin{eqnarray}
& & 
\hspace{-0.7cm}
0 = \int \mathcal{D}A \, \mathcal{D}\bar\psi \, \mathcal{D}\psi ~ \frac{\delta }{ \delta \bar\psi_\alpha (x)} ~  
\nonumber \\
& &
\hspace{-0.5cm}
\exp\left\{ i  \int d^4 x \Big[ \mathcal{L}_{QED} (A, \bar\psi, \psi) + 
   J^\mu(x)A_\mu(x) + \bar\eta(x) \, \psi(x) + \bar\psi (x)  \, \eta(x)  \Big] \right\} 
   \label{DSE0}
\end{eqnarray}
and given that
\begin{eqnarray}
  \frac{\delta S}{\delta \bar\psi_\alpha (x)} & = &
   \frac{\delta }{\delta \bar\psi_\alpha (x)} \int \, d^4 z ~ \mathcal{L}_{QED} (z) 
   \nonumber \\
   & = & \left( i \slashed{\partial} - m \right)_{\alpha\beta} \, \psi_\beta (x)  - g \, \left( \gamma^\mu \right) _{\alpha\beta} \,  A_\mu(x) \, \psi_\beta (x) \ ,
\end{eqnarray}
see  Eq.  (\ref{Eq:Lqed}), it allows to rewrite Eq. (\ref{DSE0}) as
\begin{eqnarray}
& & 
\Bigg\{ 
      \bigg[ i \slashed{\partial} - m \bigg]_{\alpha\beta} \left( \frac{\delta }{i \, \delta \bar\eta_\beta (x) }\right) ~ 
      \nonumber \\
      & & \qquad 
           - ~ g \, \left( \gamma^\mu \right)_{\alpha\beta} \left( \frac{\delta }{i \, \delta \bar\eta_\beta (x) } \right)
                                                                                 \left( \frac{\delta }{i \, \delta J^\mu (x) } \right)
      ~ + ~  \eta_\alpha (x)                                                                                  
\Bigg\} ~  Z[J, \bar\eta, \eta]  = 0 \ .
\label{DSE1}
\end{eqnarray}
The gap equation is obtained from this equation after taking its derivative with respect to $i \, \delta /  \delta \eta_\theta (y)$ resulting in
\begin{eqnarray}
& & 
 \Bigg\{ 
      i \, \bigg[ i \slashed{\partial}_x - m \bigg]_{\alpha\beta} \left( \frac{\delta^2 W [J, \bar\eta, \eta] }{ \delta \eta_\theta (y) \, \delta \bar\eta_\beta (x) }\right)  
           ~ -  ~ g \, \left( \gamma^\mu \right)_{\alpha\beta} \left( \frac{\delta^3 W [J, \bar\eta, \eta] }{\delta \eta_\theta (y) \, \delta \bar\eta_\beta (x)  \, \delta J^\mu (x) } \right)    
           \nonumber \\
           & & \qquad\qquad ~  + ~ i \, \delta_{\alpha\theta} \, \delta ( x - y )  + \cdots
\Bigg\} ~ Z[J, \bar\eta, \eta]  = 0 \ ,
\label{DSE3}
\end{eqnarray}
where $\cdots$ represent terms that vanish when the sources are set to zero. Inserting the fermion propagator 
\begin{equation}
  S_{\alpha\beta}(x -y) =  \left. \frac{\delta^2 W [J, \bar\eta, \eta] }{ \delta \bar\eta_\alpha (x)  ~ \delta \eta_\beta (y) \, } \right|_{J=\bar\eta=\eta=0} 
  \label{quarkprop}
\end{equation}
and the photon propagator
\begin{equation}
    \left.  \frac{\delta^2 W[J,\bar\eta,\eta]}{\delta J^\mu(x) ~ \delta J^\nu(y) } \right|_{J = \eta = \bar\eta = 0} ~ = ~ - ~ D_{\mu\nu} (x-y) 
    \label{Eq:PhotonProp}
\end{equation}
in Eq. (\ref{DSE3}) it becomes, after some algebra,
\begin{eqnarray}
& &
  \left[ S^{-1} (x-z) \right]_{\alpha\beta} = \left[ i \slashed{\partial}_x - m \right]_{\alpha\beta} \delta(x - z)
  \nonumber \\
  & & \qquad
     - \, i \, g \, \left( \gamma^\mu \right)_{\alpha\alpha^\prime} \, \int \, d^4y ~ \frac{ \delta^3 W }{\delta J^\mu (x) ~ \delta \eta_{\beta^\prime} (y) ~ \delta \bar\eta_{\alpha^\prime} (x) \,  }
     \left[ S^{-1} (y-z) \right]_{\beta^\prime\beta} \ .
     \label{DSE-Fermion}
\end{eqnarray}
The gap equation can be rewritten using the decomposition of the connected three-point Green function in terms of one-particle irreducible functions, see App.  
\ref{Sec:Dec} and in particular Eqs (\ref{Dec3Point}) and (\ref{Eq:Vertex_app}) , as
\begin{eqnarray}
& & 
  S^{-1} (x -  y)  = \left[ i \slashed{\partial}_x - m \right] \delta(x - y) ~
  \nonumber \\
   & & \qquad
     - ~ i \, g^2 \, \int d^4 z_1 \, d^4 z_2 ~  D_{\mu\nu} (x - z_1) ~  \Big[ \gamma^\mu ~ S(x - z_2) ~   {\Gamma}^\nu (z_2, y; z_1)  \Big]
     \label{DSE-Fermion-realspace}
\end{eqnarray}
where the photon-fermion vertex is defined as
\begin{equation} 
\left. 
\frac{\delta^3 \Gamma}{ \delta A_{c, \, \mu}  (z) \, \delta\psi_{c, \, \beta} (y) \, \delta\bar\psi_{c, \, \alpha} (x) } 
 \right|_{J=\bar\eta=\eta=0}
= 
 - \, g \, \Big( {\Gamma}^\mu \Big)_{\alpha\beta} (x, y; z) 
 \label{Eq:Vertex_app-main}
\end{equation}
that is represented in Fig. \ref{Fig:vertex} in momentum space - see also Eq. (\ref{Eq:Vert_mom}). 
The translation of  the gap equation into momentum space uses the definitions
\begin{eqnarray}
& &
   S(x -y) ~ = ~ \int \frac{d^4 p}{(2 \, \pi )^4} ~e^{- i \, p (x -y ) } ~ S(p) \ , 
   \label{Eq:SFermion_mom}\\
 & &   
   D_{\mu\nu} (x -y)  ~ = ~  \int \frac{d^4 k}{(2 \, \pi )^4} ~ e^{- i \, k (x -y ) } ~ D_{\mu\nu} (k) \ , 
   \label{Eq:Dphoton_mom} \\
 & &   
   {\Gamma}^\mu (x, y; z)   ~ = ~   \int \frac{d^4 p^\prime}{(2 \, \pi )^4} \,  \frac{d^4 p}{(2 \, \pi )^4} \, \frac{d^4 k}{(2 \, \pi )^4} ~
   e^{- i \, \left( p^\prime x + p y + k z \right) }  \nonumber \\
   & & \hspace{5cm}  ~ (2 \, \pi)^4 \, \delta( p^\prime + p + k ) ~
    {\Gamma}^\mu (p^\prime, p; k) 
    \label{Eq:Vert_mom}
\end{eqnarray}
and, after some algebra, Eq. (\ref{DSE-Fermion-realspace}) becomes
\begin{equation}
  S^{-1} (p)  = \left(  \slashed{p} - m \right)
     - \, i \, g^2 \, \int \frac{d^4 k}{( 2 \, \pi )^4} ~   D_{\mu\nu} (k) ~  \Big[ \gamma^\mu ~ S(p - k) ~   {\Gamma}^\nu (p-k, - p; k)  \Big] \ .
     \label{DSE-Fermion-momentum}
\end{equation}
The equation above is represented diagramatically in Fig. \ref{Fig:fermiongap}, where the filled blob stands for full vertices or propagators, while the tree level photon-fermion vertex is represented by a dot.
The Green functions appearing so far are bare quantities. The renormalization program for QED will be discussed later in Sec. \ref{Sec:Renor}. 

%================================================================
\subsection{The Dyson-Schwinger equation for the photon propagator}

\begin{figure}[t]
\centering
\includegraphics[width=3.5in]{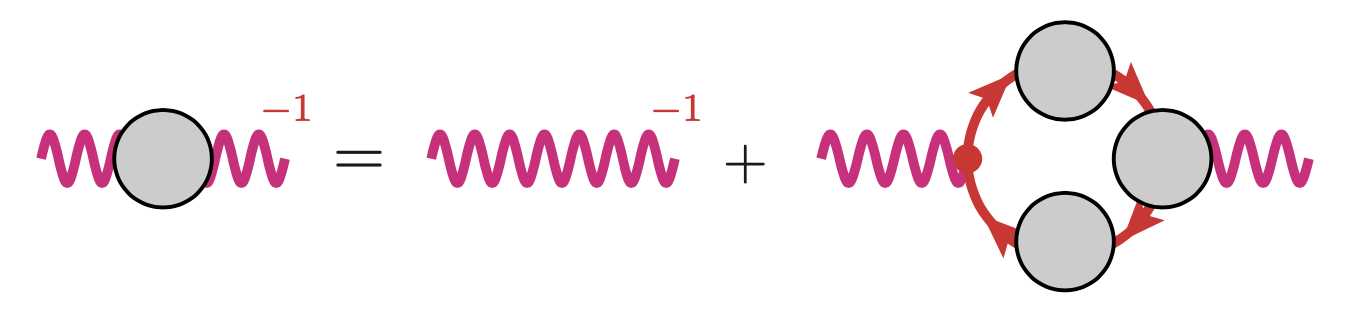} 
\caption{The photon gap equation}
\label{Fig:photongap}
\end{figure}

Similarly as in the derivation of the fermionic gap equation, the starting point to derive the DSE for the photon propagator is  the identity
\begin{eqnarray}
0 & = &
\int \mathcal{D}A \, \mathcal{D}\bar\psi \, \mathcal{D}\psi ~ \frac{\delta }{ \delta A_\mu (x)} ~  
        \nonumber \\
        & & \quad
        \exp\left\{ i  \int d^4 x \Big[ \mathcal{L}_{QED} (A, \bar\psi, \psi) +  J^\mu(x)A_\mu(x) + \bar\eta(x) \, \psi(x) + \bar\psi (x)  \, \eta(x)  \Big] \right\} \nonumber \\
 & = &
 \Bigg\{ \left[ \square_x \, g^{\mu\nu} - \left( 1 - \frac{1}{\xi} \right) \partial^\mu_x \partial^\nu_x  \right] \left( \frac{\delta}{i \, \delta J^\nu (x)} \right)
      \nonumber \\
      & & \hspace{1.5cm}
     - g \, \left( \gamma^\mu\right)_{\alpha\beta} \left( \frac{\delta }{-i \, \delta \eta_\alpha (x) } \right)  \left( \frac{\delta }{i \, \delta \bar\eta_\beta (x) } \right) 
     + J^\mu (x) 
 \Bigg\} \, Z[J, \, \eta, \, \bar\eta] \ .
 \label{DSE-photon0}
\end{eqnarray}
Performing the functional derivative with respect to $\delta / \delta J^\nu(y)$ of this equation and after setting the sources to zero one gets
\begin{eqnarray}
& &
 \left[ \square_x \, g^{\mu\mu^\prime } - \left( 1 - \frac{1}{\xi} \right) \, \partial^\mu_x \, \partial^{\mu^\prime}_x \right] \, D_{\mu^\prime\nu} (x - y ) 
 \nonumber \\
 & & \qquad\qquad
 + ~ i \, g \, \left( \gamma^\mu \right)_{\alpha\beta} \, \frac{ \delta^3 W}{\delta J^\nu(y) \, \delta\eta_\alpha (x) \, \delta\bar\eta_\beta (x)}
 ~  - ~ g^\mu_{~~ \nu}  \, \delta( x -  y ) = 0 \ .
 \label{DSE-photon1}
\end{eqnarray}
In Minkowski spacetime, the photon propagator is given by
\begin{equation}
   D_{\mu\nu} (x - y ) = \int \frac{d^4k}{(2 \, \pi)^4} ~ e^{- \, i \, k \, ( x - y ) } \underbrace{\bigg( - \, P^\perp_{\mu\nu}(k) \, D(k^2) - \frac{\xi}{k^2} \, P^L_{\mu\nu}(k) \bigg)}_{D_{\mu\nu}(k)} \ ,
   \label{PhotonPropagator}
\end{equation}
where
\begin{equation}
P^\perp_{\mu\nu}(k) = g_{\mu\nu} - \frac{k_\mu k_\nu}{k^2} \qquad\mbox{ and }\qquad
P^L_{\mu\nu}(k) = \frac{k_\mu k_\nu}{k^2}
\end{equation}
are the photon transverse and longitudinal projectors operators, respectively.
For the linear covariant gauges where $\xi \ne 0$,  Eq. (\ref{DSE-photon1}) can be multiplied by the inverse of the photon propagator and, 
after taking into account the decomposition of the three-point connected Green function (\ref{Dec3Point}), one arrives at
\begin{eqnarray}
& & 
\left( D^{-1} \right)^{\mu\nu} ( x - y )  = 
 \left[ \square_x \, g^{\mu\nu } - \left( 1 - \frac{1}{\xi} \right) \, \partial^\mu_x \, \partial^\nu_x \right] \, \delta (x - y )  
\nonumber \\
& & \hspace{1cm}
 + ~ i ~ g^2 \, \int d^4x^\prime \, d^4 x^{\prime\prime} ~ ´
            \text{Tr} \Big[ \gamma^\mu ~ S(x - x^\prime) ~ {\Gamma}^\nu (x^\prime, x^{\prime\prime}; y) ~ S(x^{\prime\prime} - x) \Big] 
             \label{DSE-photon2}
\end{eqnarray}
that in momentum space reads
\begin{eqnarray}
  \frac{P^\perp_{\mu\nu} (k)}{D(k^2)}  & = &
   k^2 \,  P^\perp_{\mu\nu}  (k) 
   \nonumber \\
   & & \quad
  - i \, g^2 \int \frac{d^4 p}{(2 \, \pi)^4} ~ \mbox{Tr} \Big[ \gamma_\mu \, S(p) \, {\Gamma}_\nu(p, -p + k; -k ) \, S(p - k ) \Big]
\end{eqnarray}
and performing the contraction of the Lorentz indices it reduces to
\begin{eqnarray}
  \frac{1}{D(k^2)}  = 
   k^2 
  - i \, \frac{g^2}{3} \,  \int \frac{d^4 p}{(2 \, \pi)^4} ~ \text{Tr} \Big[ \gamma_\mu \, S(p) \, {\Gamma}^\mu(p, -p + k; -k ) \, S(p - k ) \Big] \ .
     \label{Eq:PhotonEq_LCA}
\end{eqnarray}
This equation is represented diagramatically in Fig. \ref{Fig:photongap}.
In the derivation of Eq. (\ref{Eq:PhotonEq_LCA}) the photon propagator must be invertible, a condition that is not fulfilled in the Landau gauge.

In QED the knowledge of the fermion-photon vertex determines univocally the fermion and the photon propagator by solving Eqs.
(\ref{DSE-Fermion-momentum}) and  (\ref{Eq:PhotonEq_LCA}), respectively. Apparently, these equations do not depend on the gauge but
the vertex itself has a non-trivial dependence on the gauge fixing parameter 
that feeds the propagator equations and, in this way, they become functions of $\xi$ themselves.

%=================================================================
\subsection{A Dyson-Schwinger equation for the photon-fermion vertex \label{Sec:DSE-vertex1}}

\begin{figure}[t]
\centering
\includegraphics[width=4in]{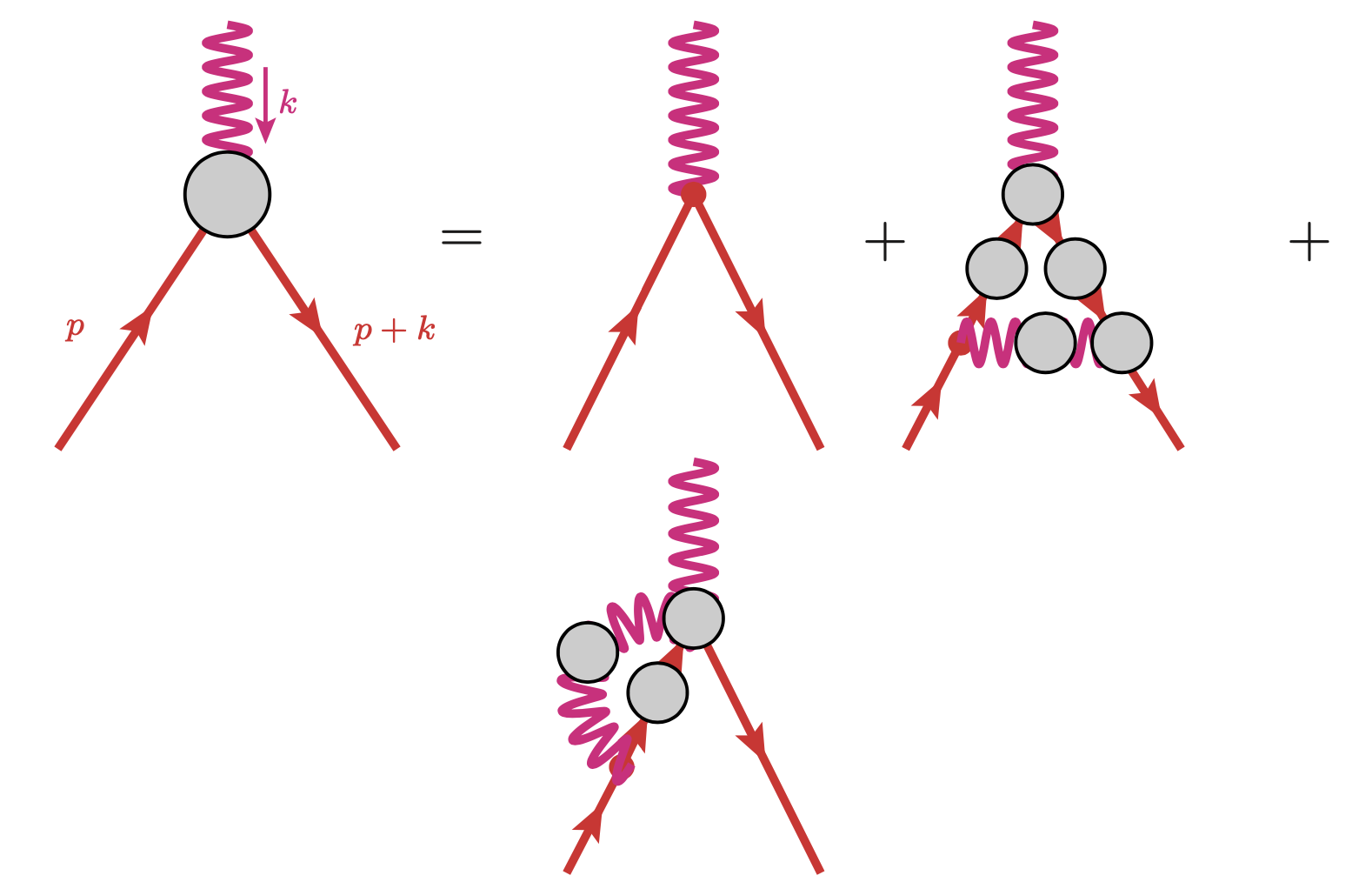} 
\caption{The Dyson-Schwinger equation for the photon-fermion vertex.}
\label{Fig:DSEvertex}
\end{figure}

As described previously, starting from Eq. (\ref{DSE1}) and taking an additional functional derivative one arrives at the gap equation (\ref{DSE3}). 
A Dyson-Schwinger equation for the vertex can be derived from this last equation by evaluating its functional derivative with respect to
$\delta / \, i \, \delta J^\nu (w)$ and taking into account those terms not represented explicitly in Eq. (\ref{DSE3}). Then,  setting the sources to zero, 
the vertex equation reads
\begin{eqnarray}
   & &
    \left[ i \slashed{\partial}_x - m \right]_{\alpha\beta^\prime} \, 
            \frac{ \delta^3 W }{\delta J^\nu (w) \, \delta \eta_\beta (y)\, \delta \bar\eta_{\beta^\prime} (x) }
            \nonumber \\
            & & \qquad\qquad
    +
    i \, g \, \left( \gamma^\mu \right)_{\alpha\beta^\prime} \, 
                     \frac{ \delta^4 W }{\delta J^\nu (w) \, \delta J^\mu (x) \, \delta \eta_\beta (y) \, \delta \bar\eta_{\beta^\prime} (x)  }    
    \nonumber \\
    & & \qquad\qquad\qquad\qquad
    - ~    g \, \left( \gamma^\mu \right)_{\alpha\beta^\prime} ~ S_{\beta^\prime\beta} (x - y) ~ D_{\mu\nu} (x - w) 
      = 0   \  
      \label{VDSE}
\end{eqnarray}
with the fermion propagator defined in Eq. (\ref{quarkprop}) and the photon propagator  settled  in Eq. (\ref{Eq:PhotonProp}).
Note that besides the propagators, the equation requires also the knowledge of the two-photon-two-fermion connected Green function. 
The decomposition of the Green functions
 that appear in (\ref{VDSE}) in terms of one-particle irreducible functions is discussed in App. \ref{Sec:Dec}; see,
in particular, Eqs (\ref{Decomp3cV0}), (\ref{Dec3Point}), (\ref{Dec3Point-mom})  and the definition (\ref{Eq:Vertex_app}) for the decomposition of the three-point function
and  Eqs (\ref{Dec4Point}), (\ref{Dec4Point-mom}) and the definitions (\ref{V_2bosons_2fermions}), (\ref{V_3bosons}) for the decomposition of the four-point function.
In momentum space the Dyson-Schwinger equation for the vertex is given by
\begin{eqnarray}
& &
   {\Gamma}^\mu (p, \, -p -k; \, k)   =  \gamma^\mu  ~ + ~ i \, g^2 \, \int \frac{d^4q}{(2 \, \pi)^4} ~D_{\zeta\zeta^\prime}(q) \nonumber \\
   & & \Bigg\{
%                                \left[ 
                                \gamma^\zeta \, S(p-q) \, {\Gamma}^\mu ( p - q, \, -p -k + q; \, k) \,
                                S(p+k-q) \, {\Gamma}^{\zeta^\prime} (p +k-q, \, -p-k; q) 
%                                \right] 
%\qquad 
\nonumber \\
   & & \qquad\qquad + ~
%                                \left[ 
                                \gamma^\zeta \, S(p-q) \, {\Gamma}^{\zeta^\prime\mu} (p - q, \, -p-k; \, q , \, k) 
%                                \right]                                
                                ~ \Bigg\} \ .
                                \label{Eq:DSE1-MOM}
\end{eqnarray}
In the derivation of the last expression the gap equation (\ref{DSE-Fermion-momentum}) was used to simplify the final result.
Furthermore, the original Dyson-Schwinger equation includes a boson propagator that needs to be canceled out to arrive at Eq. (\ref{Eq:DSE1-MOM}).
We recall that there is no inverse for the photon propagator in the Landau gauge and, therefore, the cancelation of the propagator is possible only for the linear
covariant gauges. The diagramatic representation of Eq. (\ref{Eq:DSE1-MOM}) is given in Fig. \ref{Fig:DSEvertex}.

%----------------------------------------------------

%================================================================
%================================================================
\section{Ward-Takahashi Identities \label{Sec:WTI}}

For QED the Ward-Takahashi identities can be derived by taking the variation of the generating functional (\ref{Eq:Geradores}) 
with respect to an infinitesimal gauge transformation (\ref{Eq:GaugeTransf}) that results in 
\begin{eqnarray}
& & 
 \frac{1}{g \, \xi} \, \square_x \, \partial^\mu_x \, \left( \frac{\delta Z[J, \bar\eta, \eta] }{i \, \delta J^\mu(x)} \right)
 + \frac{1}{g} \, \left( \partial^\mu_x \, J_\mu (x) \right) \, Z[J, \bar\eta, \eta]  
 \nonumber \\
 & & \qquad\qquad
 +  ~  \bar\eta_\alpha (x) \,  \left( \frac{\delta Z[J, \bar\eta, \eta] }{\delta\bar\eta_\alpha (x)} \right)
 + \left( \frac{\delta Z[J, \bar\eta, \eta] }{\delta\eta_\alpha (x)} \right) \, \eta_\alpha (x) = 0 \ .
 \label{Ward1}
\end{eqnarray}
By differentiating this equation and after setting all the sources to zero, one can arrive at the usual Ward-Takahashi identity for the fermion-photon vertex
\begin{eqnarray}
& & 
 \frac{1}{g \, \xi} \, \square_x \, \partial^\mu_x \, \left( \frac{\delta^3  Z}{i \, \delta J^\mu(x) \, \delta\eta_\beta (z) \, \delta\bar\eta_\zeta (y) } \right)
 \nonumber \\
 & & \qquad
 + \delta(x- y)   \left( \frac{\delta^2 Z }{\delta\eta_\beta (z) \, \delta\bar\eta_\zeta (x) }  \right)
 - \delta(x- z)   \left( \frac{\delta^2 Z }{\delta\eta_\beta (x) \, \delta\bar\eta_\zeta (y) }  \right) = 0
  \label{Ward2}
\end{eqnarray}
and, after further differentiation, to the Ward-Takahashi identity for the two-photon-two-fermion vertex
\begin{eqnarray}
& & 
 \frac{1}{g \, \xi} \, \square_x \, \partial^\mu_x \, 
           \left( \frac{\delta^4  Z}{ \delta J^\nu(w) \, i \, \delta J^\mu(x) \, \delta\eta_\beta (z) \, \delta\bar\eta_\zeta (y) } \right) 
           \nonumber \\
           & & \qquad\quad
                      + \frac{1}{g} ~ \bigg( \partial^x_\nu \, \delta(x - w) \bigg) ~  \left( \frac{\delta^2 Z }{\delta\eta_\beta (z) \, \delta\bar\eta_\zeta (y) }  \right) 
  \nonumber \\
 & & \qquad\qquad\qquad\quad
 + ~ \delta(x- y)   \left( \frac{\delta^3 Z }{\delta J^\nu (w) \, \delta\eta_\beta (z) \, \delta\bar\eta_\zeta (x) }  \right)
  \nonumber \\
 & & \qquad\qquad\qquad\qquad\qquad\quad
 - \delta(x- z)   \left( \frac{\delta^3  Z }{\delta J^\nu (w) \, \delta\eta_\beta (x) \, \delta\bar\eta_\zeta (y) }  \right) = 0  \ .
  \label{Ward3}
\end{eqnarray}

The identities given in Eqs (\ref{Ward2}) and (\ref{Ward3}) can be written in terms of one-particle irreducible Green functions
and in momentum space following the usual rules, see (\ref{Eq:SFermion_mom}) - (\ref{Eq:Vert_mom}) for example, and taking
into account the results and definitions given in App. \ref{Sec:Dec}. Indeed, after some straightforward algebra Eq. (\ref{Ward2}) becomes
\begin{equation}
   k_\mu  \, {\Gamma}^\mu (p, \, - p - k; \, k) = S^{-1}(p+k) \,  - \, S^{-1} (p) \ ,
    \label{Ward:photon-fermion-vertex}
\end{equation}   
while Eq.  (\ref{Ward3}) results in
\begin{equation}
  k_\mu \, \Gamma^{\mu\nu} (p, \, -p-k-q; \, k, \, q)  ~ = ~ \Gamma^\nu ( p, \, - p - q; \, q )  ~ -   ~\Gamma^{\nu} ( p+k, \, -p-k-q; \, q)\ . 
        \label{Ward:2photon2fermion}
\end{equation}
Details on the derivation of the WTI in momentum space can be found in App. \ref{Sec:WTI-Mom}.
In \cite{Scherer:1996ux,Eichmann:2012mp} the Ward-Takahashi identity for the two-photon-two-fermion vertex is also given.
Furthermore, by the contraction of Eq. (\ref{Ward:2photon2fermion}) with  momentum $q$ one arrives at the ``scalar'' WTI
\begin{eqnarray}
  k_\mu \, q_\nu \, \Gamma^{\mu\nu} (p, \, -p-k-q; \, k, \, q)  & = &  S^{-1}(p+k) ~ + ~ S^{-1}(p+q) 
  \nonumber \\
  & & \qquad
  ~ - ~ S^{-1}(p+k+q) ~ - ~S^{-1}(p) \, .  
        \label{Ward:2photon2fermionScalar}
\end{eqnarray}
that is symmetric under interchange of $k \leftrightarrow q$ and the r.h.s. requires only the knowledge of the fermion propagator.

For massless QED in 1+1 spacetime dimensions, also named the Schwinger model, the Ward-Takahashi identity for the two-photon-two-fermion 
vertex was deduced and explored in \cite{Radozycki:1998pf}. The WTI's for QED, see Eq. (\ref{Ward:2photon2fermion}), and for the
Schwinger model are related to their Lagrangian densities that differ mainly in their dimensions. Moreover, the derivation of the WTIs
are independent of the fermion mass and the Ward-Takahashi identity for QED and for the Schwinger model
are identical. We  call the reader attention that in \cite{Radozycki:1998pf} the ``vertex function''  $\Gamma^{\mu\nu}$ 
does not have the same meaning as in the present manuscript. Indeed, in \cite{Radozycki:1998pf} the function  $\Gamma^{\mu\nu}$ 
represents the four point connect Green function, see their App., while here it represent the two-photon-two-fermion
one-particle irreducible Green function. Once this difference is taken into account, the WTI become identical.

%================================================================
%================================================================
\section{Solving the Ward-Takahashi identities \label{Sec:SolWTI-II}}

The Ward-Takahashi identities translate into the Green functions of QED the constraints due the gauge symmetry. 
Then, any realistic description of QED should comply with the WTI and, therefore, they play a major role in modelling the vertices. 
The constraints due to the WTI do not apply to the tensor structures that describe the full vertices but constraint only
its longitudinal components, relative to the photon momenta.

%================================================================
%================================================================
\subsection{The Ward-Takahashi identity for the photon-fermion vertex \label{Sec:SolWTI-vertex}}

The WTI for the photon-fermion vertex given in Eq. (\ref{Ward:photon-fermion-vertex}) has been studied long ago \cite{Ball:1980ay} and is 
commonly used to fix the longitudinal part of the vertex. 
In the following lines, we review the Ball-Chiu construction for the longitudinal part of the photon-fermion vertex also to fix the notation. 

The photon-fermion vertex \cite{Ball:1980ay,Oliveira:2018ukh}  can be written
in terms of a longitudinal $\Gamma_L$ and transverse $\Gamma_T$ components, relative to the photon momentum, as
\begin{equation}
 \Gamma^\mu (p_2, \, p_1; \, p_3)  = \Gamma^\mu_L (p_2, \, p_1; \, p_3) +  \Gamma^\mu_T (p_2, \, p_1; \, p_3) \ ,
\end{equation}
where $p_2$ is the incoming fermion momentum, $-p_1$ is the outgoing fermion momentum and $p_3$ the incoming photon momenta. The
momentum are such that $p_1 + p_2 + p_3 = 0$. The Feynman diagram representing the vertex is given in Fig. \ref{Fig:vertex}.
It follows from the definition that 
\begin{equation}
 p_{3 \, _\mu } \,  \Gamma^\mu_T (p_2, \, p_1; \, p_3)  = 0 
 \end{equation}
 and the WTI (\ref{Ward:photon-fermion-vertex}) reads
\begin{equation}
   p_{3 \, _\mu}  \, {\Gamma}^\mu (p_2, p_1; \, p_3) = p_{3 \, _\mu}  \, {\Gamma}^\mu_L (p_2, p_1; \, p_3) =  S^{-1}(-p_1) \,  - \, S^{-1} (p_2) \ .
    \label{WTI-vertex}
\end{equation}
By writing the longitudinal vertex using the Ball-Chiu basis of operators \cite{Ball:1980ay}, i.e. setting
\begin{eqnarray}
 & & 
   \Gamma^\mu_L (p_2, \, p_1; \, p_3) = \lambda_1 (p^2_1, \, p^2_2, \, p^2_3) \,  \gamma^\mu  ~ + ~ \lambda_2 (p^2_1, \, p^2_2, \, p^2_3)  \, \Big( \slashed{p}_1 - \slashed{p}_2 \Big) \Big( p_1 - p_2 \Big)^\mu
   \nonumber \\
   & & \qquad\quad %\hspace{3cm}
   + ~ \lambda_3 (p^2_1, \, p^2_2, \, p^2_3)  ~ \Big( p_1 - p_2 \Big)^\mu
   ~ + ~ \lambda_4 (p^2_1, \, p^2_2, \, p^2_3)  ~ \sigma^{\mu\nu}  \,  \Big( p_1 - p_2 \Big)_\nu \ ,
   \label{Eq:Long-Vertex}
\end{eqnarray}
where $\lambda_i$ are Lorentz scalar form factors, $\sigma_{\mu\nu} = [ \gamma_\mu \, , \, \gamma_\nu ]/2$, then, writing the 
inverse of the fermion propagator as
\begin{equation}
  S^{-1}(p) = A(p^2) \, \slashed{p} - B(p^2) \ ,
  \label{Eq:InverseFermProp}
\end{equation}
one can solve the WTI  and compute the various form factors in terms of $A$ and $B$ that are given by
\begin{eqnarray}
  \lambda_1 (p^2_1, \, p^2_2, \, p^2_3) & = & \frac{1}{2} \bigg( A\big( p^2_1 \big)  + A\big(p^2_2\big) \bigg)  \ ,    \label{EQ:L1} \\
  \lambda_2 (p^2_1, \, p^2_2, \, p^2_3) & = & \frac{1}{2 \, \left( p^2_1 - p^2_2 \right)}   \bigg( A\big( p^2_1 \big)  -  A\big(p^2_2\big) \bigg) \ , \label{EQ:L2} \\
  \lambda_3 (p^2_1, \, p^2_2, \, p^2_3) & = & \frac{1}{ p^2_1 - p^2_2 }   \bigg( B\big(p^2_1\big) - B\big( p^2_2 \big)    \bigg) \ ,   \label{EQ:L3} \\
  \lambda_4 (p^2_1, \, p^2_2, \, p^2_3) & = & 0 \ .  \label{EQ:L4}
\end{eqnarray}
Such as parameterisation of the vertex is known in the literature as the Ball-Chiu vertex and will be used to describe the longitudinal
part of the photon-fermioin vertex here.  The non-vanishing form factors of the Ball-Chiu vertex, i.e. $\lambda_i$ with $i = 1, 2, 3$,
are symmetric functions under exchange of the first two arguments
as required by charge conjugation, see e.g. \cite{Davydychev:2000rt}. 
Moreover, the form factors that solve the WTI have no kinematic singularities even in the limit where 
$p_1 \rightarrow p_2$. Indeed, setting $p_1 = p_2 + \delta$, assuming that $A$ and $B$ are smooth functions of the momentum, then after
expanding all functions in powers of $\delta$ and taking the limit $\delta \rightarrow 0$ it follows that $\lambda_2$ and $\lambda_3$ 
are proportional to derivatives of $A$ and $B$ and, therefore, these form factors, being smooth functions of momentum,
are not singular for $p_1 = p_2$.

%================================================================
%================================================================
\subsection{The Ward-Takahashi identity for the two-photon-two-fermion vertex \label{Sec:SolWTI-2P2F}}

Before discussing the solution of the WTI for the two-photon-two-fermion irreducible Green function, let us complete the description of photon-fermion
vertex, whose knowledge is required by WTI itself. The longitudinal component of the photon fermion vertex was described in
(\ref{Eq:Long-Vertex}) using the Ball-Chiu basis and the $\lambda_i$ are taken from the WTI. Similarly as for the longitudinal part of the vertex,
let as introduce a basis of operators to describe $\Gamma_T$. The transverse component of the vertex requires eight form factors, i.e.
\begin{equation}
 \Gamma_{T \, \mu} (p_2, \, p_1; p_3) =  \sum^8_{i=1} \tau_i (p^2_1, \, p^2_2, \, p^2_3) \, T^{(i)}_\mu (p_1, \, p_2, \,  p_3) \ .
 \label{Eq:photon-fermion_vertex-trans}
\end{equation}
Here, we take the K{\i}z{\i}lersu-Reenders-Pennington \cite{Kizilersu:1995iz} basis of operators 
\begin{eqnarray}
T^{(1)}_\mu (p_1, \, p_2, \,  p_3) & = & p_{1 \, _\mu} \big( p_2 \cdot p_3 \big) - p_{2 \, _\mu}  \big( p_1 \cdot p_3 \big)  \ ,   \label{TensorBasis-T1}  \\
T^{(2)}_\mu (p_1, \, p_2, \,  p_3) & = & - \, T^{(1)}_\mu (p_1, \, p_2, \,  p_3) ~ \big( \slashed{p}_1 - \slashed{p}_2 \big)  \ , \label{TensorBasis-T2}  \\
T^{(3)}_\mu (p_1, \, p_2, \,  p_3) & = & p^2_3 \, \gamma_\mu - p_{3 \, _\mu} \, \slashed{p}_3  \ , \label{TensorBasis-T3}  \\
T^{(4)}_\mu (p_1, \, p_2, \,  p_3) & = & T^{(1)}_\mu (p_1, \, p_2, \,  p_3) ~ \sigma_{\alpha\beta} \, p^\alpha_1 \,  p^\beta_2 \ , \label{TensorBasis-T4} \\
T^{(5)}_\mu (p_1, \, p_2, \,  p_3) & = & \sigma_{\mu\nu} \, p^\nu_3 \ , \label{TensorBasis-T5}  \\
T^{(6)}_\mu (p_1, \, p_2, \,  p_3) & = & \gamma_\mu \big( p^2_1 - p^2_2 \big) + \big( p_{1} - p_{2} \big)_\mu \, \slashed{p}_3 \ , \label{TensorBasis-T6}  \\
T^{(7)}_\mu (p_1, \, p_2, \,  p_3) & = &  - \, \frac{1}{2} \, \big( p^2_1 - p^2_2 \big) \, \big[ \gamma_\mu \,  \big( \slashed{p}_1 - \slashed{p}_2 \big)  - \big( p_{1} - p_{2} \big)_\mu\big]  
\nonumber \\
& & \qquad\qquad
               - \big( p_{1} - p_{2} \big)_\mu ~ \sigma_{\alpha\beta} \, p^\alpha_1 \,  p^\beta_2 \ , \label{TensorBasis-T7}  \\
T^{(8)}_\mu (p_1, \, p_2, \,  p_3) & = &  - \, \gamma_\mu \, \sigma_{\alpha\beta} \, p^\alpha_1 \,  p^\beta_2  \, + \, p_{1 \, _\mu} \slashed{p}_2 \, - \,  p_{2 \, _\mu} \slashed{p}_1  \ ,\label{TensorBasis-T8} 
\end{eqnarray}
that verify 
\begin{equation}
p^\mu_3 \, T^{(i)}_\mu (p_1, \, p_2, \,  p_3) = 0 \ .
\end{equation}
The transverse form factors are free of kinematical singularities. Having chosen a basis of operators, computing the vertex is equivalent 
to determine the form factors $\tau_i$. The computation of the $\tau_i$'s can be done in several ways, for example in perturbation theory
\cite{Kizilersu:1995iz} or solving the DSE (\ref{Eq:DSE1-MOM}). To proceed with the computation of a solution of the two-photon-two-fermion WTI
it will be assumed that the full photon-fermion vertex is known.

Let us turn our attention to the Ward-Takahashi idendity for the two-photon-two-fermion vertex, see Eq. (\ref{Ward:2photon2fermion}),
and discuss its solutions starting by studying the soft photon limit where the photon momenta $q$ vanishes. 
For this kinematics the r.h.s. of the WTI is determined by photon-fermion vertices with a vanishing photon momentum $q = 0$
that requires only its longitudinal component. Inserting the longitudinal components, see Eqs (\ref{EQ:L1}) to  (\ref{EQ:L4}), 
on the r.h.s. of the WTI,  it is straightforward to show that
\begin{eqnarray}
& & 
\Gamma^{\mu\nu}_L (p, \, -p-k; \, k, \, 0)  =  
g^{\mu\nu} ~ \bigg[ 2 \, \lambda_3 \left( (p + k )^2, \, (p+k)^2, \, 0 \right)  \bigg]  \nonumber \\
& & \quad + ~
 \frac{k^\mu \,p^\nu}{k^2}  ~  \bigg[ 2 \, \left(  \lambda_3 \left( (p + k )^2, \, (p+k)^2, \, 0 \right) - \lambda_3 \left( p^2, \, p^2, \, 0 \right)\right) \bigg] \nonumber \\
 & & \quad + ~
 \frac{k^\mu \, \gamma^\nu}{k^2} ~ \bigg[ \lambda_1 \left( p^2, \, p^2, \, 0 \right) - \lambda_1 \left( (p + k )^2, \, (p+k)^2, \, 0 \right)   \bigg] \nonumber \\
 & & 
\quad  + ~ \slashed{k} ~ 
              \bigg(  ~ g^{\mu\nu} ~ + ~ \frac{k^\mu \, p^\nu}{k^2}  \bigg) ~ \bigg[ -4 \, \lambda_2 \left( (p + k )^2, \, (p+k)^2, \, 0 \right)  \bigg] \nonumber \\
 & & 
  \quad + ~ \slashed{p} ~
  \Bigg\{  ~ g^{\mu\nu} \bigg[ -4 \, \lambda_2 \left( (p + k )^2, \, (p+k)^2, \, 0 \right) \bigg] \nonumber \\
       & & \qquad\quad ~ + ~ \frac{k^\mu \, p^\nu}{k^2}  \bigg[ 4 \, \left( \lambda_2 \left( p^2, \, p^2, \, 0 \right) - \lambda_2 \left( (p + k )^2, \, (p+k)^2, \, 0 \right) \right)  \bigg]
           \Bigg\} \ .
           \label{Eq:WTI-Sol-SoftPhoton}
\end{eqnarray}
Note that we wrote $\Gamma^{\mu\nu}_L$ instead of $\Gamma^{\mu\nu}$ as the 
WTI is blind to any component orthogonal to the photon momenta $k$.
For a vanishing photon momenta $\lambda_2$ and $\lambda_3$ are
proportional to derivatives of $A$ and $B$, respectively, and if these functions are essentially constant then the above solution of the WTI can be
approximated by
\begin{equation}
\Gamma^{\mu\nu}_L (p, \, -p-k; \, k, \, 0)  = 
 \frac{k^\mu \, \gamma^\nu}{k^2} ~ \bigg[ A( p^2) \,  - \,  A( (p + k )^2)   \bigg]  \ .
           \label{Eq:WTI-Sol-SoftPhoton-Approx}
\end{equation}
As for the photon-fermion vertex, the solutions of the WTI in the soft photon are determined by the inverse fermion propagator form factors $A$ and $B$,
and its derivatives.

Let's proceed and discuss the low momenta approximation to (\ref{Ward:2photon2fermion}). 
Our understanding of the low momenta approximation is a linear approximation in $q$ and $k$ to the r.h.s. of the WTI, 
disregarding all higher order contributions in any of the photon momenta. Then, the longitudinal components of the photon-fermion vertex to
the l.h.s. of the WTI read
\begin{eqnarray}
& &
   k_\mu \Bigg\{ 
    2 \, B^\prime (p^2) \, g^{\mu\nu}  ~ + ~ 4 \, B^{\prime\prime}(p^2) \, p^\mu \, p^\nu ~
    - ~ 2 \, A^\prime(p^2) \, \left( p^\mu \, \gamma^\nu \, + \, p^\nu \,\gamma^\mu \right)  
    \nonumber \\
    & & \qquad\qquad\qquad
   ~ -  ~ 2 \, A^\prime(p^2) \, \slashed{p} \, g^{\mu\nu} 
     ~ - ~ 4 \, A^{\prime\prime}(p^2) \, \slashed{p} \, p^\mu \, p^\nu  \Bigg\} 
\end{eqnarray}
and, therefore, it follows that
\begin{eqnarray}
& & 
\hspace{-0.5cm}
\Gamma^{\mu\nu}_L (p, \, -p-k-q; \, k, \, q)  = 
     2 \, B^\prime (p^2) \, g^{\mu\nu}  ~ + ~ 4 \, B^{\prime\prime}(p^2) \, p^\mu \, p^\nu
      \nonumber \\
     & & ~ 
   - ~ 2 \, A^\prime(p^2) \, \left( p^\mu \, \gamma^\nu \, + \, p^\nu \,\gamma^\mu \right)  ~ -  ~ 2 \, A^\prime(p^2) \, \slashed{p} \, g^{\mu\nu} 
     ~ - ~ 4 \, A^{\prime\prime}(p^2) \, \slashed{p} \, p^\mu \, p^\nu \ ,
     \label{Eq:WTI-Sol-Linear}
\end{eqnarray}
modulo possible contributions coming from the transverse part of the photon-fermion vertex. 
$A^\prime$ and $A^{\prime\prime}$ stand for the first and second derivatives of $A$ with respect to its argument,
with $B^\prime$ and $B^{\prime\prime}$ having a similar meaning for the function $B$.
Now, let us discuss the contribution of the transverse operators $T^{(i)}_\mu$ given in Eqs (\ref{TensorBasis-T1}) to (\ref{TensorBasis-T8}). 
The operators $T^{(i)}_\mu$ are either linear of quadratic in $q$ and $k$ and, therefore, they are associated with higher order contributions
in the photon momenta to the r.h.s. of the WTI. Since the linear contribution to the r.h.s of the WTI vanishes, then
the expression in Eq. (\ref{Eq:WTI-Sol-Linear}) solves the WTI within the linear approximation. 
This solution is symmetric under exchange of $\mu \leftrightarrow \nu$, as required by Bose symmetry, 
and calls for the first and second derivatives of the fermion propagator form factors, evaluated at the incoming fermion momentum.
At large $p^2$ one expects to recover the perturbative result to be a valid approximation. At large momenta, 
$A$ and $B$ are essentially constants and, therefore, give sub-leading contributions to $\Gamma^{\mu\nu}_L$, i.e. in the ultraviolet regime
$\Gamma^{\mu\nu}_L \sim 0$.

The analysis of the linear approximation to the WTI identifies an hierarchy of tensors that are required to described the longitudinal
part of the two-photon-two-fermion vertex that are associated with the longitudinal photon-fermion vertex.
This suggest a possible construction for a partial tensor basis to write $\Gamma^{\mu\nu}_L$. Indeed, by looking
at the contribution of the longitudinal photon-fermion vertex to the WTI and including only the linear, in the photon momenta, transverse operators (after
imposing Bose symmetry), one can arrives at
\begin{eqnarray}
& &
\Gamma^{\mu\nu}(p, \, -p-k-q; \, k, \, q)_L  = 
\Gamma_0 \, g^{\mu\nu} ~ + ~ \Gamma_1 \, p^\mu \, p^\nu
~ + ~ \Gamma_2 \, \left(  p^\mu \, \gamma^\nu \, + \, p^\nu \, \gamma^\mu \right)
\nonumber \\
& & 
~ + ~ \Gamma_3 \, \slashed{p} \, g^{\mu \nu}
~ + ~ \Gamma_4 \, \slashed{p} \, p^\mu \, p^\nu 
~ + ~ \Gamma_5 \, \left(  \frac{k^\mu}{k^2} \sigma^{\nu\alpha} q_\alpha  \, + \, \frac{q^\nu}{q^2} \sigma^{\mu\alpha} k_\alpha\right) \nonumber \\
 & & \quad
 + ~ \Gamma_6 \, \Bigg(  \frac{k^\mu}{k^2} \Big( \,  (pq) \gamma^\nu \, - \,  p^\nu \, \slashed{q} \,  \Big)  \, + \, 
                                       \frac{q^\nu}{q^2} \Big( \,  (pk) \gamma^\mu \, - \,  p^\mu \, \slashed{k} \,  \Big) \Bigg)  \nonumber \\
 & & \quad
 + ~ \Gamma_7 \, \Bigg(  \frac{k^\mu}{k^2} \Big[ \, (pq) \Big( \gamma^\nu \, \slashed{p} - p^\nu \Big) + p^\nu \, \sigma_{\alpha\beta} \, p^\alpha q^\beta \Big]
      \nonumber \\
      & & \hspace{3cm} 
  \, + \, \frac{q^\nu}{q^2} \Big[ \, (pk) \Big( \gamma^\mu \, \slashed{p} - p^\mu \Big) + p^\mu \, \sigma_{\alpha\beta} \, p^\alpha k^\beta \Big] \Bigg)
  \nonumber \\
 & & \quad
 + ~ \Gamma_8 \, \Bigg(  \frac{k^\mu}{k^2} \Big[  \gamma^\nu \sigma_{\alpha\beta} \, q^\alpha p^\beta + p^\nu \, \slashed{q} - q^\nu \, \slashed{p} \Big] 
       \nonumber \\
      & & \hspace{3cm} 
  \, + \, \frac{q^\nu}{q^2} \Big[  \gamma^\mu \sigma_{\alpha\beta} \, k^\alpha p^\beta + p^\mu \, \slashed{k} - k^\nu \, \slashed{p} \Big]  \Bigg) \ .
  \label{Eq:MinimalBasis-2P2F}
\end{eqnarray}
The discussion of this particular basis and its relation with other solutions will be postponed until Sec. \ref{Sec:2P2F-Tensor-Basis}.

Let us start to discuss the solution of the WTI (\ref{Ward:2photon2fermion}) by looking at its contraction with $q$,
that results in the scalar version of the WTI given in (\ref{Ward:2photon2fermionScalar}). 
The scalar WTI is symmetric under interchange of photon momenta and suggests the following ``solution''
\begin{eqnarray}
& & 
  \Gamma^{\mu\nu}_L (p, \, -p-k-q; \, k, \, q)   
  \nonumber \\
  & & =   
      \frac{k^\mu \, q^\nu}{k^2 q^2} \bigg(  k_\mu \Gamma^\mu( p, \, -p-k; k)  - k_\mu \Gamma^\mu( p+q, \, -p-k-q; k )   \bigg) \nonumber \\
  &  & = 
  \frac{k^\mu \, q^\nu}{k^2 q^2} \bigg( S^{-1}(p+k) ~ + ~ S^{-1}(p+q) ~ - ~ S^{-1}(p+k+q) ~ - ~S^{-1}(p) \bigg) \! .
        \label{Ward:2photon2fermionScalarXX}
\end{eqnarray}
The term in parenthesis on r.h.s. vanish when $k$ and/or $q$ approach zero and it opens the possibility of having a ``solution''
with no kinematical singularities at zero momenta. The above expression does not solve the WTI exactly. Going back to
the WTI (\ref{Ward:2photon2fermion}) and consider the ansatz
\begin{eqnarray}
& & 
   \Gamma^{\mu\nu}_L (p, \, -p-k-q; \, k, \, q)   =   \frac{k^\mu}{k^2} \bigg( \Gamma^\nu ( p, \, - p - q; \, q )  ~ -   ~\Gamma^{\nu} ( p+k, \, -p-k-q; \, q) \bigg)  \nonumber \\
   & & \qquad\qquad + ~  \frac{q^\nu}{q^2} \bigg( \Gamma^\mu ( p, \, - p - k; \, k )  ~ -   ~\Gamma^{\mu} ( p+q, \, -p-k-q; \, k) \bigg) \nonumber \\
   & & \qquad\qquad + ~P^{\perp \, \mu\nu}(k) ~ \widetilde{A}(k,q) + ~ P^{L \, \mu\nu}(k) ~  \widetilde{B}(k,q) \nonumber \\
   & &  \qquad\qquad + ~P^{\perp \, \mu\nu}(q) ~ \widetilde{A}(q,k) + ~ P^{L \, \mu\nu}(q) ~  \widetilde{B}(q,k) \ ,
   \label{Sol:WTITwoPhotonTwoFermion1}
\end{eqnarray}         
where $\widetilde{A}$ and $\widetilde{B}$ are functions of the photon momenta. This ansatz solves the two-photon-two-fermion WTI 
if, for all $k$ and $q$,
\begin{eqnarray}
   & &   \frac{q^\nu}{q^2} \bigg(      S^{-1}(p+k) ~ + ~ S^{-1}(p+q) ~ - ~  S^{-1}(p) ~ - ~ S^{-1}(p+k+q)    \bigg) 
   \nonumber \\
   & & ~ + 
   \bigg( k^\nu - \frac{(k \cdot q) }{q^2} q^\nu \bigg)  \, \widetilde{A}(q,k) 
   ~ + ~ k^\nu \, \widetilde{B}(k,q) ~ + ~ \frac{(k \cdot q)}{q^2} \,  q^\nu  \, \widetilde{B}(q,k) = 0  .
\end{eqnarray}
To arrive at this later expression, the  photon-fermion WTI (\ref{Ward:photon-fermion-vertex}) was used to rewrite the equation.
A general solution of this equation can be found  if the functions $\widetilde{A}$ and $\widetilde{B}$ are symmetric under interchange of $k$ and $q$.
In this case, it follows that
\begin{eqnarray}
& & 
  \widetilde{B}(k,q)  =  - \, \widetilde{A}(q,k) \ ,    \label{Sol:WTITwoPhotonTwoFermion2} \\
  & & 
  \widetilde{A}(q,k)  = 
  \nonumber \\
  & & ~ \frac{1}{2 \, (k \cdot q)}  \bigg( 
        S^{-1}(p+k) ~ + ~ S^{-1}(p+q) ~ - ~  S^{-1}(p) ~ - ~ S^{-1}(p+k+q)    \bigg)  \nonumber \\
         & = & \frac{1}{2 \, (k q)} \Bigg\{
                \bigg[ A\left( (p+k)^2\right) + A\left( (p+q)^2\right) - A\left( (p+k+q)^2\right) - A\left( p^2\right) \bigg] ~\slashed{p} \nonumber \\
                & & \quad\quad + ~  \bigg[ A\left( (p+k)^2\right)  - A\left( (p+k+q)^2\right) \bigg] ~\slashed{k} \nonumber \\
                & & \quad\quad + ~  \bigg[ A\left( (p+q)^2\right)  - A\left( (p+k+q)^2\right) \bigg] ~\slashed{q} \nonumber \\                
                & & \quad\quad +  ~ B\left( (p+k+q)^2\right) + B\left( p^2\right) -  B\left( (p+k)^2\right) - B\left( (p+q)^2\right) \Bigg\}  .
           \label{Sol:WTITwoPhotonTwoFermion3}
 \end{eqnarray}

The solution of the WTI   summarized in Eqs  (\ref{Sol:WTITwoPhotonTwoFermion1}), (\ref{Sol:WTITwoPhotonTwoFermion2}) and 
(\ref{Sol:WTITwoPhotonTwoFermion3}) writes $\Gamma^{\mu\nu}_L$ in terms of the photon-fermion vertex and 
the form factors $A$ and $B$ that describe the fermion propagator.
Furthermore, the expression in Eq. (\ref{Sol:WTITwoPhotonTwoFermion3}) is free of kinematical singularities and, therefore,
$\Gamma^{\mu\nu}_L$ is also free of kinematical singularities. Moreover,
for the perturbative tree level solution of the theory, where $A = 1$, $B = m$ and the photon-fermion vertex reads $\Gamma^\mu = \gamma^\mu$,
the longitudinal component of the two-photon-two-fermion irreducible vertex vanishes as expected, and
a non-vanishing $\Gamma^{\mu\nu}_L$ within the perturbative solution of QED can only be associated with the loop corrections.

%We conclude this section with a note on the solution of the WTI for the two-photon-two-fermion one particle irreducible diagram for the
%Schwinger model discussed in \cite{Radozycki:1998pf}. As mentioned at the end of Sec. \ref{Sec:WTI}, the corresponding WTI does
%not match exactly with the QED identity. However, comparing both solutions one can find similar Dirac algebraic tensor structures, 
%with the QED solution calling for extra terms not seen in the Schwinger model.

%======================================================
%======================================================
\section{The Dyson-Schwinger equation for the two-photon-two-fermion  irreducible vertex \label{Sec:VertexfromWTI}}

\begin{figure}[t]
\centering
\includegraphics[width=2in]{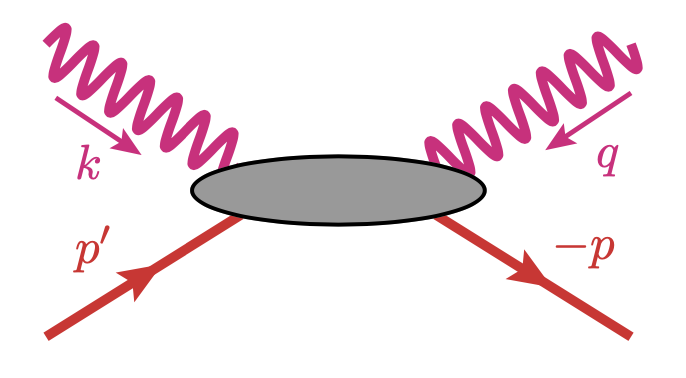} 
\caption{The two-photon-two-fermion one-particle irreducible vertex: see Eq. (\ref{Eq:TwoPhotonTwoFermionVertexDef}) for defintions.}
\label{Fig:TwoPhotonTwoFermion}
\end{figure}

As discussed in the previous section, the longitudinal part of the two-photon-two-fermion vertex is determined by the corresponding WTI. However,
the full $\Gamma^{\mu\nu}$ certainly goes beyond $\Gamma^{\mu\nu}_L$ and one needs to be able to compute it. Here, we consider the DySon-Schwinger
equation for this irreducible vertex that, in principle, can allow to access the remaining components.

The one-particle irreducible two-photon-two-fermion vertex represented in Fig. \ref{Fig:TwoPhotonTwoFermion} and defined in Eq. (\ref{V_2bosons_2fermions}),
\begin{eqnarray}
   g^2 \, {\Gamma}^{\mu\nu}_{\alpha\beta} (x, y; z,w)   & = & - ~
   \frac{\delta^4\Gamma}{\delta A_{cl, \, \mu} (z) ~ \delta A_{cl, \, \nu} (w) ~ \delta\psi_{cl, \, \beta} (y) ~ \delta\bar\psi_{cl, \, \alpha} (x)} 
         \nonumber \\
         & = &  g^2 
   \int \frac{d^4 p^\prime}{(2 \, \pi )^4}   \frac{d^4 p}{(2 \, \pi )^4}  \frac{d^4 k}{(2 \, \pi )^4}  \frac{d^4 q}{(2 \, \pi )^4} ~
   e^{- i \, \left( p^\prime x + p y + k z + q w \right) }  \nonumber \\
   & & \hspace{1.3cm} \, (2 \, \pi)^4 ~ \delta( p^\prime + p + k + q ) ~
    {\Gamma}^{\mu\nu}_{\alpha\beta}  (p^\prime, \, p; \, k, \, q) \ , 
    \label{Eq:TwoPhotonTwoFermionVertexDef}
\end{eqnarray}
is symmetric under interchange of the photon indices. As the photon-fermion irreducible vertex, 
${\Gamma}^{\mu\nu}$ can be decomposed into a  longitudinal ${\Gamma}^{\mu\nu}_L$
and  an orthogonal ${\Gamma}^{\mu\nu}_T$ component,  relative to the photon momenta, i.e. in momentum space
\begin{eqnarray}
& & 
{\Gamma}^{\mu\nu} = {\Gamma}^{\mu\nu}_L + {\Gamma}^{\mu\nu}_T
\nonumber \\
& &
\qquad \mbox{ with } \qquad
k_\mu \,  {\Gamma}^{\mu\nu}_T(p^\prime, \, p; \, k, \, q) ~ = ~  q_\nu \, {\Gamma}^{\mu\nu}_T(p^\prime, \, p; \, k, \, q)  ~ = ~ 0   .
\end{eqnarray}
The Ward-Takahashi identity (\ref{Ward:2photon2fermion}) constrains only the longitudinal part of $\Gamma^{\mu\nu}$.
The solution of the WTI discussed in Sec. \ref{Sec:SolWTI}, 
see Eqs (\ref{Sol:WTITwoPhotonTwoFermion1}) and (\ref{Sol:WTITwoPhotonTwoFermion3}), writes
 $\Gamma^{\mu\nu}_L$ in terms of the photon-fermion vertex and the fermion propagator functions $A$ and $B$.
 
\begin{figure}[t]
\centering
\includegraphics[width=4in]{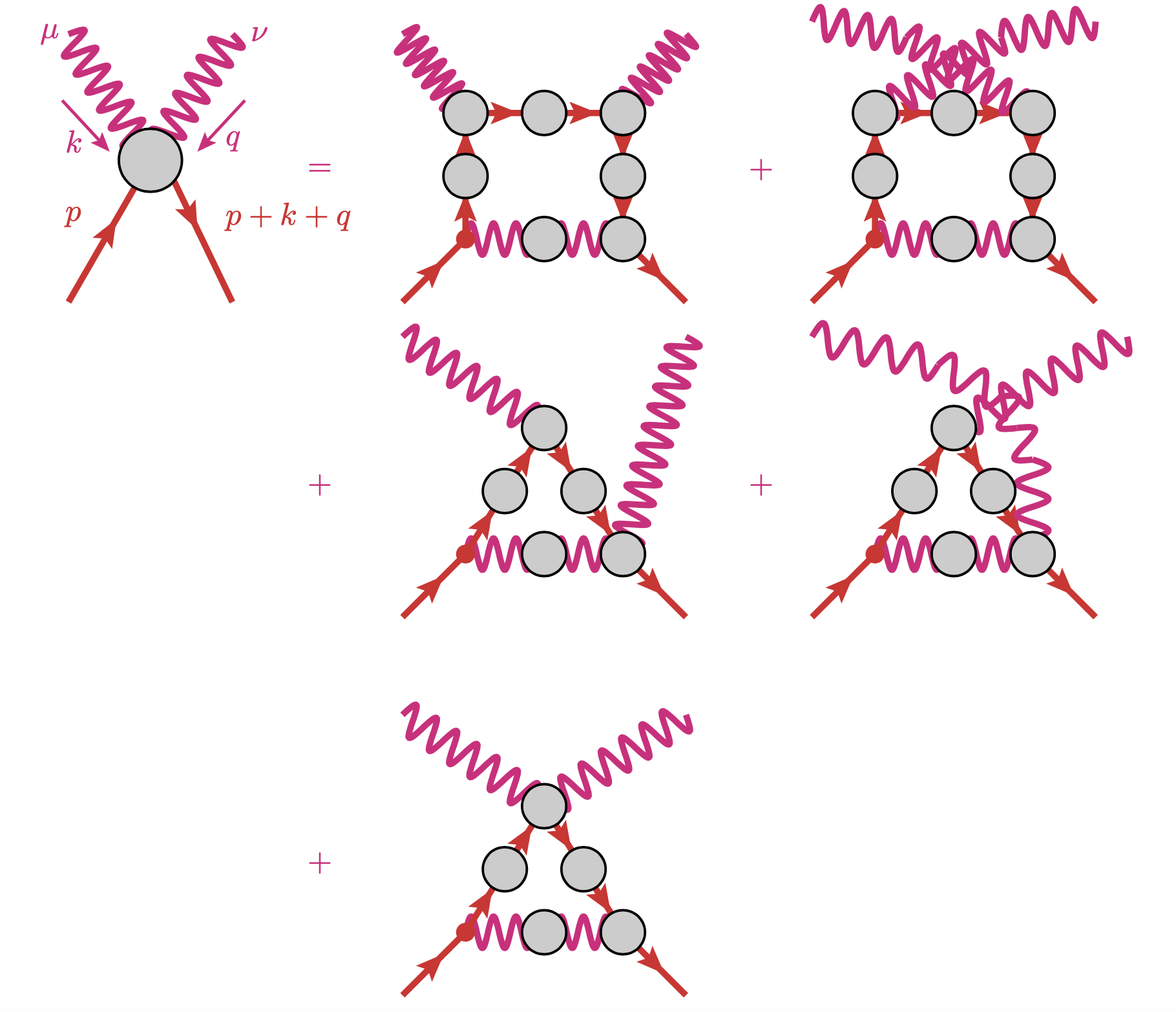} 
\caption{The truncated Dyson-Schwinger equation for the two-photon-two-photon-fermion one-particle irreducible vertex as a function of the one-particle irreducible vertices and full propagators. Recall that only vertices with a number of external legs smaller than four were considered.}
\label{Fig:DSEtwophoton}
\end{figure}

 The Dyson-Schwinger equation for the two-photon-two-fermion Green function $\Gamma^{\mu\nu}$ 
 can  be obtained in the usual way after taking various  functional derivatives of Eq. (\ref{DSE1}), and after setting the sources to zero it reads
\begin{eqnarray}
& &
  0 = \left( i \, \slashed{\partial}_x - m \right)_{\alpha\beta} 
               \frac{\delta^4 W}{\delta J^\rho (z) \, \delta J^\nu (w) \, \delta\eta_\gamma (y) \, \delta\bar\eta_\beta (x)}  \nonumber \\
   & &
   \hspace{1cm}  + ~ i ~ g \, \left( \gamma^\mu \right)_{\alpha\beta}  \,
                   \frac{\delta^5 W}{\delta J^\rho (z) \, \delta J^\nu (w) \,  \delta J ^\mu (x) \, \delta\eta_\gamma (y) \, \delta\bar\eta_\beta (x)} \nonumber \\
   & &
    \hspace{1.5cm}  - ~  g \, \left( \gamma^\mu \right)_{\alpha\beta}  \, 
                          \frac{\delta^3 W}{ \delta J ^\nu (w) \, \delta\eta_\gamma (y) \, \delta\bar\eta_\beta (x)}  \, 
                          \frac{\delta^2 W}{\delta J^\rho (z) \, \delta J^\mu (x)}  \nonumber \\
   & &
    \hspace{2cm} 
   - ~  g \, \left( \gamma^\mu \right)_{\alpha\beta}  \, 
                          \frac{\delta^3 W}{ \delta J ^\rho (z) \, \delta\eta_\gamma (y) \, \delta\bar\eta_\beta (x)}  \,
                          \frac{\delta^2 W}{\delta J^\nu (w) \, \delta J^\mu (x)} \nonumber \\
   & &
    \hspace{2.5cm} 
   - ~  g \, \left( \gamma^\mu \right)_{\alpha\beta}  \, 
                          \frac{\delta^2 W}{ \delta\eta_\gamma (y) \, \delta\bar\eta_\beta (x)}  \,
                          \frac{\delta^3 W}{\delta J ^\rho (z) \, \delta J^\nu (w) \, \delta J^\mu (x)}  \nonumber \\
& &
  + ~  \Bigg(  i \, \left( i \, \slashed{\partial}_x - m \right)_{\alpha\beta} \, 
                                                  \frac{\delta^2 W}{\delta\eta_\gamma (y) \, \delta\bar\eta_\beta (x)} ~ - ~  g \, \left( \gamma^\mu \right)_{\alpha\beta}  \, 
                          \frac{\delta^3 W}{ \delta J ^\mu (x) \, \delta\eta_\gamma (y) \, \delta\bar\eta_\beta (x)}   
                          \nonumber \\
                          & & \hspace{4.7cm} ~ +  ~i \, \delta_{\alpha\gamma} \, \delta (x-y)  \Bigg) \, \frac{\delta^2 W}{\delta J ^\rho (z) \, \delta J^\nu (w)}  
                          \ .
  \label{Eq:Nova_2fotoes_2fermions-0}
\end{eqnarray}
The term in parenthesis,  see the last line, is the fermion gap equation (\ref{DSE3}) and, therefore, it vanishes.
The last term in Eq. (\ref{Eq:Nova_2fotoes_2fermions-0}) before the one proportional to the gap equation is proportional to the three-photon 
one-particle irreducible vertex and it vanishes for QED (Furry's theorem). 
It turns out that the DSE for the two-photon-two-fermion  reads
\begin{eqnarray}
& &
   \left( i \, \slashed{\partial}_x - m \right)_{\alpha\beta} 
               \frac{\delta^4 W}{\delta J^\rho (z) \, \delta J^\nu (w) \, \delta\eta_\gamma (y) \, \delta\bar\eta_\beta (x)}  \nonumber \\
   & &
   \hspace{0.6cm}  + ~ i ~ g \, \left( \gamma^\mu \right)_{\alpha\beta}  \,
                   \frac{\delta^5 W}{\delta J^\rho (z) \, \delta J^\nu (w) \,  \delta J ^\mu (x) \, \delta\eta_\gamma (y) \, \delta\bar\eta_\beta (x)} \nonumber \\
   & &
    \hspace{1.2cm}  - ~  g \, \left( \gamma^\mu \right)_{\alpha\beta}  \, 
                          \frac{\delta^3 W}{ \delta J ^\nu (w) \, \delta\eta_\gamma (y) \, \delta\bar\eta_\beta (x)}  \, 
                          \frac{\delta^2 W}{\delta J^\rho (z) \, \delta J^\mu (x)}  \nonumber \\
   & &
    \hspace{1.8cm} 
   - ~  g \, \left( \gamma^\mu \right)_{\alpha\beta}  \, 
                          \frac{\delta^3 W}{ \delta J ^\rho (z) \, \delta\eta_\gamma (y) \, \delta\bar\eta_\beta (x)}  \,
                          \frac{\delta^2 W}{\delta J^\nu (w) \, \delta J^\mu (x)} = 0 \ .
  \label{Eq:Nova_2fotoes_2fermions}
\end{eqnarray}
To arrive at a solution of Eq. (\ref{Eq:Nova_2fotoes_2fermions}) for the two-photon-two-fermion vertex, all the connected Green functions have to be
written 
in terms of one-particle irreducible vertices.
The second term of this equation requires a five-point connected Green function, whose 
decomposition in terms of one-particle irreducible vertices can be achieved following the type of calculations discussed in App \ref{Sec:Dec}. 
For the five-point function in (\ref{Eq:Nova_2fotoes_2fermions}), after performing the necessary functional derivatives 
of the generating functionals, after some algebra and after setting the sources to zero one arrives at
\begin{small}
\begin{eqnarray}
& & 
  \frac{\delta^5 W}{\delta J_\rho (s) \, \delta J_\nu (w) \, \delta J_\mu (z) \, \delta\eta_\beta (y) \, \delta\bar\eta_\alpha  (x) \,  } ~ = ~\nonumber \\
  & & 
  = ~ 
  \int  d^4v_1 \, d^4v_2 \, d^4v_3 \, d^4v_4  \, d^4v_5 ~ 
                 \propagator{J_\mu(z)}{J^{\mu^\prime} (v_1)} ~ \propagator{J_\nu(w)}{J^{\nu^\prime} (v_2)} ~ \propagator{J_\rho(s)}{J^{\rho^\prime} (v_3)} 
                 \nonumber \\
                 & & \hspace{3cm}
                  \propagator{\eta_{\alpha^\prime} (v_4)}{\bar\eta_\alpha (x)} ~\propagator{\eta_{\beta} (y)}{\bar\eta_{\beta^\prime} (v_5)}
                   \nonumber \\
                  & & \hspace{3cm}
                  \oneparticle{5}{\delta A_{c, \, \rho^\prime}(v_3) ~\delta A_{c, \, \nu^\prime} (v_2) ~\delta A_{c, \, \mu^\prime} (v_1) 
                                        ~ \delta\psi_{c, \, \beta^\prime} (v_5) ~ \delta\bar\psi_{c, \, \alpha^\prime} (v_4) }  \nonumber \\
 & &                          
 + ~    \int  d^4v_1 \, d^4v_2 \, d^4v_3  ~ 
                 \propagator{J_\rho(s)}{J^{\rho^\prime} (v_1)} ~  \propagator{\eta_{\alpha^\prime} (v_2)}{\bar\eta_\alpha (x)} 
                 ~ \connect{4}{\delta J_\nu (w) ~ \delta J_\mu (z) ~ \delta \eta_\beta (y) ~ \delta\bar\eta_{\beta^\prime} (v_3)}
                  \nonumber \\
                  & & \hspace{3cm}
                  \oneparticle{3}{\delta A_{c, \, \rho^\prime}(v_1) ~ \delta\psi_{c, \, \beta^\prime} (v_3) ~ \delta\bar\psi_{c, \, \alpha^\prime} (v_2) }  \nonumber \\
 & &                          
 + ~    \int  d^4v_1 \, d^4v_2 \, d^4v_3  ~ 
                 \propagator{J_\nu (w)}{J^{\nu^\prime} (v_1)} ~  \propagator{\eta_{\alpha^\prime} (v_2)}{\bar\eta_\alpha (x)} 
                 ~ \connect{4}{\delta J_\rho (s) ~ \delta J_\mu (z) ~ \delta \eta_\beta (y) ~ \delta\bar\eta_{\beta^\prime} (v_3)}
                  \nonumber \\
                  & & \hspace{3cm}
                  \oneparticle{3}{\delta A_{c, \, \nu^\prime}(v_1) ~ \delta\psi_{c, \, \beta^\prime} (v_3) ~ \delta\bar\psi_{c, \, \alpha^\prime} (v_2) }  \nonumber \\
 & &                          
 + ~    \int  d^4v_1 \, d^4v_2 \, d^4v_3  ~ 
                 \propagator{J_\mu (z)}{J^{\mu^\prime} (v_1)} ~  \propagator{\eta_{\alpha^\prime} (v_2)}{\bar\eta_\alpha (x)} 
                 ~ \connect{4}{\delta J_\rho (s) ~ \delta J_\nu (w) ~ \delta \eta_\beta (y) ~ \delta\bar\eta_{\beta^\prime} (v_3)}
                  \nonumber \\
                  & & \hspace{3cm}
                  \oneparticle{3}{\delta A_{c, \, \mu^\prime}(v_1) ~ \delta\psi_{c, \, \beta^\prime} (v_3) ~ \delta\bar\psi_{c, \, \alpha^\prime} (v_2) }   \nonumber \\
 & &                          
 + ~    \int  d^4v_1 \, d^4v_2 \, d^4v_3 ~ 
                 \propagator{\eta_{\alpha^\prime} (v_1)}{\bar\eta_\alpha (x)} ~  \propagator{\eta_{\beta} (y)}{\bar\eta_{\beta^\prime} (v_2)} 
                 ~ \connect{4}{\delta J_\rho (s) ~ \delta J_\nu (w) ~ \delta J_\mu (z) ~ \delta J^{\zeta^\prime} (v_3)}
                  \nonumber \\
                  & & \hspace{3cm}
                  \oneparticle{3}{\delta A_{c, \, \zeta^\prime}(v_3) ~ \delta\psi_{c, \, \beta^\prime} (v_2) ~ \delta\bar\psi_{c, \, \alpha^\prime} (v_1) }   \nonumber \\
 & &                          
 + ~    \int  d^4v_1 \, d^4v_2 \, d^4v_3 \, d^4v_4  ~ 
                 \propagator{J_\rho (s)}{J^{\rho^\prime} (v_1)} ~ \propagator{J_\nu (w)}{J^{\nu^\prime} (v_2)} ~
                 \propagator{\eta_{\alpha^\prime} (v_3)}{\bar\eta_\alpha (x)} \nonumber \\
                 & & 
                 \hspace{1.4cm} 
                 \connect{3}{\delta J_\mu (z) ~ \delta\eta_{\beta} (y) ~ \delta\bar\eta_{\beta^\prime} (v_4)} ~
                  \oneparticle{4}{\delta A_{c, \, \rho^\prime}(v_1) ~ \delta A_{c, \, \nu^\prime}(v_2) ~ \delta\psi_{c, \, \beta^\prime} (v_4) 
                                                      ~ \delta\bar\psi_{c, \, \alpha^\prime} (v_3) }   \nonumber \\
 & &                          
 + ~    \int  d^4v_1 \, d^4v_2 \, d^4v_3 \, d^4v_4  ~ 
                 \propagator{J_\rho (s)}{J^{\rho^\prime} (v_1)} ~ \propagator{J_\mu (z)}{J^{\mu^\prime} (v_2)} ~
                 \propagator{\eta_{\alpha^\prime} (v_3)}{\bar\eta_\alpha (x)} \nonumber \\
                 & & 
                 \hspace{1.4cm} 
                 \connect{3}{\delta J_\nu (w) ~ \delta\eta_{\beta} (y) ~ \delta\bar\eta_{\beta^\prime} (v_4)} ~
                  \oneparticle{4}{\delta A_{c, \, \rho^\prime}(v_1) ~ \delta A_{c, \, \mu^\prime}(v_2) ~ \delta\psi_{c, \, \beta^\prime} (v_4) 
                                                      ~ \delta\bar\psi_{c, \, \alpha^\prime} (v_3) } \nonumber \\
 & &                          
 + ~    \int  d^4v_1 \, d^4v_2 \, d^4v_3 \, d^4v_4  ~ 
                 \propagator{J_\nu (w)}{J^{\nu^\prime} (v_1)} ~ \propagator{J_\mu (z)}{J^{\mu^\prime} (v_2)} ~
                 \propagator{\eta_{\alpha^\prime} (v_3)}{\bar\eta_\alpha (x)} \nonumber \\
                 & & 
                 \hspace{1.4cm} 
                 \connect{3}{\delta J_\rho (s) ~ \delta\eta_{\beta} (y) ~ \delta\bar\eta_{\beta^\prime} (v_4)} ~
                  \oneparticle{4}{\delta A_{c, \, \nu^\prime}(v_1) ~ \delta A_{c, \, \mu^\prime}(v_2) ~ \delta\psi_{c, \, \beta^\prime} (v_4) 
                                                      ~ \delta\bar\psi_{c, \, \alpha^\prime} (v_3) }  \ . 
             \label{Eq:Dec5points-exact} 
\end{eqnarray}
\end{small}
The decomposition given in Eq. (\ref{Eq:Dec5points-exact}) for the five-point Green function is symmetric under the interchange of any pair of the
bosonic degrees of freedom as required by Bose symmetry. The decomposition of the five-point Green function requires further Green functions not
considered so far. In order to arrive at a manageable equation and to avoid the introduction of new vertices, we consider a truncation of
Eq. (\ref{Eq:Dec5points-exact}) and ignore the terms that are proportional to the  one-particle irreducible five-point Green function (first term on r.h.s.)
and the four-photon irreducible  diagrams that appear in the decomposition of the four-photon connected Green function. 
By disregarding the four-photon irreducible vertices it turns out that the contribution of the connected four-photon Green function, that is 
a sum of a term containing the irreducible four-photon vertex and terms proportional to the three-photon irreducible vertices, that vanish in QED (Furry's theorem), 
is not taken into consideration within the truncation considered. 
By taking into account the definition of the photon and fermion propagators,
the five-point Green function given in Eq. (\ref{Eq:Dec5points-exact}) 
can be simplified using the definitions given in Eqs (\ref{Eq:Vertex_app}) and (\ref{V_2bosons_2fermions}) for the one-particle irreducible Green functions.
In momentum space the above equation becomes
\begin{eqnarray}
& & 
  \frac{\delta^5 W}{\delta J^\rho (s) \, \delta J^\nu (w) \, \delta J^\mu (z) \, \delta\eta_\beta (y) \, \delta\bar\eta_\alpha  (x) \,  } ~ = ~ \nonumber \\
  & & 
  = ~ g^3 ~ \int \frac{d^4 k_1}{(2 \, \pi)^4} \, \frac{d^4 k_2}{(2 \, \pi)^4} \, \frac{d^4 k_3}{(2 \, \pi)^4}  \, \frac{d^4 p}{(2 \, \pi)^4}  ~
     e^{- \, i \, k_1 z - \, i \, k_2 w - \, i \, k_3 s - \, i \, p \, x + \, i \, ( p + k_1 + k_2 + k_3 ) y} \nonumber \\
     & & \qquad\qquad
    D_{\rho\rho^{\prime}}(k_3) ~ D_{\nu\nu^{\prime}}(k_2) ~ D_{\mu\mu^{\prime}}(k_1) ~\Bigg\{ ~S (p)  ~ \bigg[  \nonumber \\
       & &% \hspace{1cm}
        ~
         \Gamma^{\rho^\prime}(p, \, - p-k_3; \, k_3) \, S(p+k_3) \, \Gamma^{\nu^\prime}(p + k_3, \, -p-k_2-k_3; k_2) \, S(p+k_2+k_3) \, \nonumber \\
                                                                                  & & \hspace{4cm}
                                                                                                       \Gamma^{\mu^\prime}(p+k_2+k_3, \, p-k_1-k_2-k_3; k_1) \nonumber \\
        & & %\hspace{1cm}  
         + ~
         \Gamma^{\rho^\prime}(p, \, -p-k_3; \, k_3) \, S(p+k_3) \, \Gamma^{\mu^\prime}(p+k_3, \, -p-k_1-k_3; k_1) \, S(p+k_1+k_3) \, \nonumber \\
                                                                                  & & \hspace{4cm}
                                                                                                       \Gamma^{\nu^\prime}(p+k_1+k_3, \, -p-k_1-k_2-k_3; k_2) \nonumber \\
        & & %\hspace{1cm}  
         + ~
         \Gamma^{\nu^\prime}(p, \, -p-k_2; \, k_2) \, S(p+k_2) \, \Gamma^{\rho^\prime}(p+k_2, \, -p-k_2-k_3; k_3) \, S(p+k_2+k_3) \,  \nonumber \\
                                                                                  & & \hspace{4cm}
                                                                                                       \Gamma^{\mu^\prime}(p+k_2+k_3, \, -p-k_1-k_2-k_3; k_1) \nonumber \\
        & & %\hspace{1cm}   
        + ~
         \Gamma^{\nu^\prime}(p, \, -p-k_2; \, k_2) \, S(p+k_2) \, \Gamma^{\mu^\prime}(p+k_2, \, -p-k_1-k_2; k_1) \, S(p+k_1+k_2) \, \nonumber \\
                                                                                  & & \hspace{4cm}
                                                                                                       \Gamma^{\rho^\prime}(p+k_1+k_2, \, -p-k_1-k_2-k_3; k_3) \nonumber \\
        & & %\hspace{1cm}   
        + ~
         \Gamma^{\mu^\prime}(p, \, -p-k_1; \, k_1) \, S(p+k_1) \, \Gamma^{\rho^\prime}(p+k_1, \, -p-k_1-k_3; k_3) \, S(p+k_1+k_3) \,  \nonumber \\
                                                                                  & & \hspace{4cm}
                                                                                                       \Gamma^{\nu^\prime}(p+k_1+k_3, \, -p-k_1-k_2-k_3; k_2) \nonumber \\
        & & %\hspace{1cm}  
         + ~
         \Gamma^{\mu^\prime}(p, \, -p-k_1; \, k_1) \, S(p+k_1) \, \Gamma^{\nu^\prime}(p+k_1, \, -p-k_1-k_2; k_2) \, S(p+k_1+k_2) \, \nonumber \\
                                                                                  & & \hspace{4cm}
                                                                                                       \Gamma^{\rho^\prime}(p+k_1+k_2, \, -p-k_1-k_2-k_3; k_3)   \nonumber \\
        %===============================
    & & %\hspace{1cm}    
    + ~
        \Gamma^{\rho^\prime}(p, \, -p-k_3; \, k_3) \, S(p+k_3) \, \Gamma^{\nu^\prime\mu^\prime} (p+k_3, \, -p-k_1-k_2-k_3; \, k_2, \, k_1) \nonumber \\
        & & %\hspace{1cm}   
        + ~
         \Gamma^{\nu^\prime}(p, \, -p-k_2; \, k_2) \, S(p+k_2) \, \Gamma^{\rho^\prime\mu^\prime}(p+k_2, \, -p-k_1-k_2-k_3; \, k_3, \, k_1) \nonumber \\
        & & %\hspace{1cm}   
        + ~
         \Gamma^{\mu^\prime}(p, \, -p-k_1; \, k_1) \, S(p+k_1) \, \Gamma^{\rho^\prime\nu^\prime}(p+k_1, \, -p-k_1-k_2-k_3; \, k_3, \, k_2) \nonumber \\
         & & %\hspace{1cm}  
          + ~
         \Gamma^{\rho^\prime\nu^\prime}(p, \, -p-k_2-k_3; \, k_3, \, k_2)  \, S(p+k_2+k_3) \,
               \nonumber \\
               & & \hspace{4cm} \Gamma^{\mu^\prime}(p+k_2+k_3, \, -p-k_1-k_2-k_3; \, k_1) \nonumber \\
         & & %\hspace{1cm}   
         + ~
         \Gamma^{\rho^\prime\mu^\prime}(p, \, -p-k_1-k_3; \, k_3, \, k_1)  \, S(p+k_1+k_3) \, 
                        \nonumber \\
               & & \hspace{4cm} \Gamma^{\nu^\prime}(p+k_1+k_3, \, -p-k_1-k_2-k_3; \, k_2) \nonumber \\
         & & %\hspace{1cm}   
         + ~
         \Gamma^{\nu^\prime\mu^\prime}(p, \, -p-k_1-k_2; \, k_2, \, k_1)  \, S(p+k_1+k_2) \, 
                        \nonumber \\
               & & \hspace{4cm}\Gamma^{\rho^\prime}(p+k_1+k_2, \, -p-k_1-k_2-k_3; \, k_3) ~\bigg]
         \nonumber \\
          & & %\hspace{1cm}
        ~S(p+k_1+k_2+k_3) ~ \Bigg\}_{\alpha\beta}  \ .
             \label{Eq:Dec5points-mom}
\end{eqnarray}
Proceeding, as usual, starting from Eq. (\ref{Eq:Nova_2fotoes_2fermions}), 
doing the decomposition of the Green functions in terms of one-particle irreducible vertices and propagators, performing the necessary Fourier transformations and doing the required inversions, one arrives at the truncated Dyson-Schwinger equation for two-photon-two-fermion 
one-particle irreducible vertex in momentum space
\begin{eqnarray}
& &
\Gamma^{\mu\nu} (p, \, -p-k-q; \, k, \, q)  ~ = \nonumber \\
& &  \quad =
  ~ Z^\mu (p,k) ~  \bigg[ S(p+k) ~ \Gamma^\nu ( p+k, \, -p-k-q; \, q) \bigg] ~  \nonumber \\
& & \qquad
+ ~ Z^\nu (p, q) ~ \bigg[ S(p+q) ~ \Gamma^\mu ( p+q, \, -p-k-q; \, k) \bigg] \nonumber \\
& & \qquad - ~  \Gamma^\mu (p, \, -p-k; \, k) ~S(p+k) ~  \Gamma^\nu (p+k, \, -p-k-q; \, q)  \nonumber \\
& & \qquad - ~ \Gamma^\nu (p, \, -p-	q; \, q) ~S(p+q) ~  \Gamma^\mu (p+q, \, -p-k-q; \, k)  \nonumber \\
& & \quad\quad+ \, i \, g^2 ~\int \frac{d^4 w}{(2 \, \pi)^4} ~ D_{\zeta\zeta^\prime}(w) ~\gamma^\zeta ~S(p-w)  \nonumber \\
        & &  \bigg\{ 
                     ~ \Gamma^\mu (p-w, \, -p-k+w; \, k) ~ S(p+k-w) ~ 
                     \nonumber \\
                     & & \hspace{2cm} \Gamma^{\nu} (p+k-w, \, -p-k-q+w; \, q)  \nonumber \\
                     & & \hspace{2cm} S(p+k+q-w) ~ \Gamma^{\zeta^\prime} (p+k+q-w, \, -p-k-q; \, w)        \nonumber \\
        & &  ~ + ~
                     ~\Gamma^\nu (p-w, \, -p-q+w; \, q) ~ S(p+q-w) ~ 
                     \nonumber \\
                     & & \hspace{2cm} \Gamma^{\mu} (p+q-w, \, -p-k-q+w; \, k)  \nonumber \\
                     & & \hspace{2cm} S(p+k+q-w) ~ \Gamma^{\zeta^\prime} (p+k+q-w, \, -p-k-q; \, w)        \nonumber \\
        & &  ~ + ~
                     \Gamma^\mu (p-w, \, -p-k+w; \, k) ~ S(p+k-w) ~ 
                     \nonumber \\
                     & & \hspace{2cm} \Gamma^{\zeta^\prime} (p+k-w, \, -p-k; \, w)  \nonumber \\
                     & & \hspace{2cm} S(p+k) ~ \Gamma^{\nu} (p+k, \, -p-k-q; \, q)      \nonumber \\
        & & ~ + ~
                     ~\Gamma^\nu (p-w, \, -p-q+w; \, q) ~ S(p+q-w) ~ 
                     \nonumber \\
                     & & \hspace{2cm} \Gamma^{\zeta^\prime} (p+q-w, \, -p-q; \, w)  \nonumber \\
                     & & \hspace{2cm} S(p+q) ~ \Gamma^{\mu} (p+q, \, -p-k-q; \, k)        \nonumber \\                     
        & & ~ + ~
                     ~\Gamma^\mu (p-w, \, -p-k+w; \, k) ~ S(p+k-w) ~  
                     \nonumber \\
                     & & \hspace{2cm} \Gamma^{\nu\zeta^\prime} (p+k-w, \, -p-k-q; \, q, \, w)        \nonumber \\                     
        & &  ~ + ~
                     ~\Gamma^\nu (p-w, \, -p-q+w; \, q) ~ S(p+q-w) ~ 
                     \nonumber \\
                     & & \hspace{2cm}  \Gamma^{\mu\zeta^\prime} (p+q-w, \, -p-k-q; \, k, \, w)       \nonumber \\
        & & ~ + ~
                     ~\Gamma^{\mu\nu} (p-w, \, -p-k-q+w; \, k, \, q) ~ S(p+k+q-w) ~  
                     \nonumber \\
                     & & \hspace{2cm} \Gamma^{\zeta^\prime} (p+k+q-w, \, -p-k-q; \, w)     %  \nonumber \\
%                   & &  XXXXX
~ \bigg\}  % \nonumber \\
  \label{Eq:DSE_2photon2fermion-truncated0}
\end{eqnarray}
where
\begin{eqnarray}
& & 
   Z^\mu(p,k)  =  \gamma^\mu ~ +
   \nonumber \\
   & &
     ~ i \, g^2 \, \int \frac{d^4 w}{(2 \, \pi)^4} ~ D_{\zeta\zeta^\prime}(w) ~  \bigg[ \gamma^\zeta ~ S(p-w) ~ 
                              \Gamma^{\mu\zeta^\prime}(p -w, \, -p-k; \, k, \, w) \bigg]\ .
                              \label{Func_Z}
\end{eqnarray}
The definition of $Z^\mu$ replicates (up to a sign) the correction of the fermion self-energy
after replacing $\Gamma^\mu$ by the one-particle irreducible vertex
$\Gamma^{\mu\nu}$. 
Moreover, equation (\ref{Eq:DSE_2photon2fermion-truncated0}) is symmetric under the interchange of any pair of
bosonic indices as required by Bose statistics. The above equation can be further simplified with the help of the Dyson-Schwinger equation
for the vertex (\ref{Eq:DSE1-MOM}). Indeed, replacing in the third and fourth lines the vertex that appears on the left by 
Eq. (\ref{Eq:DSE1-MOM}), it turns out that it cancels exactly the first and the second lines of Eq. (\ref{Eq:DSE_2photon2fermion-truncated0})
and also the third and fourth lines under the integral. Then, it follows that the truncated equation determining the two-photon-two-fermion one-particle
irreducible vertex is
\begin{eqnarray}
& &
\Gamma^{\mu\nu} (p, \, -p-k-q; \, k, \, q)  ~ = \nonumber \\
& &   =
   i \, g^2 ~\int \frac{d^4 w}{(2 \, \pi)^4} ~ D_{\zeta\zeta^\prime}(w) ~\gamma^\zeta ~S(p-w)  \nonumber \\
        & & \hspace{0.5cm} \bigg\{ 
                     ~ \Gamma^\mu (p-w, \, -p-k+w; \, k) ~ S(p+k-w) ~ 
                     \nonumber \\
                     & & \hspace{3cm} \Gamma^{\nu} (p+k-w, \, -p-k-q+w; \, q)  \nonumber \\
                     & & \hspace{3cm} S(p+k+q-w) ~ \Gamma^{\zeta^\prime} (p+k+q-w, \, -p-k-q; \, w)        \nonumber \\
        & & \hspace{0.5cm} ~ + ~
                     ~\Gamma^\nu (p-w, \, -p-q+w; \, q) ~ S(p+q-w) ~ 
                     \nonumber \\
                     & & \hspace{3cm} \Gamma^{\mu} (p+q-w, \, -p-k-q+w; \, k)  \nonumber \\
                     & & \hspace{3cm} S(p+k+q-w) ~ \Gamma^{\zeta^\prime} (p+k+q-w, \, -p-k-q; \, w)        \nonumber \\
%        & & \hspace{1.5cm} ~ + ~
%                     \Gamma^\mu (p-w, \, -p-k+w; \, k) ~ S(p+k-w) ~ \Gamma^{\zeta^\prime} (p+k-w, \, -p-k; \, w)  \nonumber \\
%                     & & \hspace{4cm} S(p+k) ~ \Gamma^{\nu} (p+k, \, -p-k-q; \, q)      \nonumber \\
%        & & \hspace{1.5cm} ~ + ~
%                     ~\Gamma^\nu (p-w, \, -p-q+w; \, q) ~ S(p+q-w) ~ \Gamma^{\zeta^\prime} (p+q-w, \, -p-q; \, w)  \nonumber \\
%                     & & \hspace{4cm} S(p+q) ~ \Gamma^{\mu} (p+q, \, -p-k-q; \, k)        \nonumber \\                     
        & & \hspace{0.5cm} ~ + ~
                     ~\Gamma^\mu (p-w, \, -p-k+w; \, k) ~ S(p+k-w) ~  
                     \nonumber \\
                     & & \hspace{3cm} \Gamma^{\nu\zeta^\prime} (p+k-w, \, -p-k-q; \, q, \, w)        \nonumber \\                     
        & & \hspace{0.5cm} ~ + ~
                     ~\Gamma^\nu (p-w, \, -p-q+w; \, q) ~ S(p+q-w) ~ 
                     \nonumber \\
                     & & \hspace{3cm}  \Gamma^{\mu\zeta^\prime} (p+q-w, \, -p-k-q; \, k, \, w)       \nonumber \\
        & & \hspace{0.5cm} ~ + ~
                     ~\Gamma^{\mu\nu} (p-w, \, -p-k-q+w; \, k, \, q) ~ S(p+k+q-w) ~ 
                     \nonumber \\
                     & & \hspace{3cm}  \Gamma^{\zeta^\prime} (p+k+q-w, \, -p-k-q; \, w)     %  \nonumber \\
%                   & &  XXXXX
~ \bigg\}  % \nonumber \\
  \label{Eq:DSE_2photon2fermion-truncated}
\end{eqnarray}
whose representation in terms of Feynman diagrams is given in Fig. \ref{Fig:DSEtwophoton}.
This equation can, in principle,  be solved self-consistently as the computation of the r.h.s. requires the knowledge of
$\Gamma^{\mu\nu}$ itself.  
We end this section by calling the reader attention that the derivation of Eq. (\ref{Eq:DSE_2photon2fermion-truncated}) 
calls for the inverse of the photon propagator and of the fermion propagator that are not well defined in the Landau gauge.

%-----------------------------------PPPPPPPP
%----------------------------------------------------
%........................................PPPPPPP

%======================================================
%======================================================
\subsection{Perturbative analysis of the DSE for the two-photon-two-fermion vertex}

The Dyson-Schwinger Eq. (\ref{Eq:DSE_2photon2fermion-truncated}) is an approximation to the bare DSE for the two-photon-two-fermion irreducible 
vertex. This equation was built ignoring contributions from higher order Green functions and, therefore, the truncation can introduce unwanted 
terms that have to be removed. However, 
the diagramatic representation of the equation, see Fig. \ref{Fig:DSEtwophoton}, and the perturbative analysis 
of Eq. (\ref{Eq:DSE_2photon2fermion-truncated}) shows that, for the truncation considered, this is not the case.

In perturbation theory all quantities are written as series in the coupling constant.  To proceed let us introduce the  notation
for the photon-fermion vertex
\begin{equation}
\Gamma^\mu (p_2, \, p_1; \, p_3) = \gamma^\mu + \Delta\Gamma^\mu (p_2, \, p_1; \, p_3) \ ,
\label{Eq:GammaPert}
\end{equation}
where $\Delta\Gamma^\mu$ represents the corrections to the tree-level vertex. In the perturbative solution of QED, $\Delta\Gamma^\mu$ starts at order $g^2$. 
Indeed, in the lowest order in the coupling constant, see Eq. (\ref{Eq:DSE1-MOM}), it is given by
\begin{eqnarray}
& & 
   \Delta{\Gamma}^\mu (p, \, -p -k; \, k)  ~  = ~   i \, g^2 \, \int \frac{d^4w}{(2 \, \pi)^4} ~D_{\zeta\zeta^\prime}(w) ~
   \nonumber \\
   & & \hspace{4.7cm}
                                \bigg[ 
                                \gamma^\zeta \, S(p-w) \, \gamma^\mu ~ S(p+k-w) ~ \gamma^{\zeta^\prime} 
                                \bigg] \ .
                                \label{Eq:DSE1-MOM-pert} 
\end{eqnarray}
Inserting the definition (\ref{Eq:GammaPert}) into Eq. (\ref{Eq:DSE_2photon2fermion-truncated}), 
it follows that the lowest order contribution to $\Gamma^{\mu\nu}$ is of order $g^2$. Thus, in lowest order in the coupling constant
one has
\begin{eqnarray}
& & 
  \Gamma^{\mu\nu} ( p, \, -p-k-q; \, k, \, q)  =   i \, g^2 \, \int \frac{d^4 w}{(2 \, \pi)}  ~    D_{\zeta\zeta^\prime}(w)  
%   \gamma^\zeta ~ S(p -w) ~ \gamma^\nu ~S(p+q-w) ~\gamma^{\zeta^\prime} ~S(p+q) ~\gamma^\mu \nonumber \\
%  & &  \qquad\qquad\qquad
%   + ~
%      \gamma^\zeta ~ S(p -w) ~ \gamma^\mu ~S(p+k-w) ~\gamma^{\zeta^\prime} ~S(p+k) ~\gamma^\nu \nonumber \\
%  & &  \qquad\qquad\qquad
%   + ~
      \nonumber \\
      & & \hspace{2cm} 
      \gamma^\zeta  ~ S(p -w) \times
      \nonumber \\
      & & \hspace{2.5cm}  \times \bigg\{ ~  \gamma^\nu  ~ S(p+q-w) ~ \gamma^\mu %~S(p+k+q-w) ~\gamma^{\zeta^\prime} 
%      \nonumber \\
%  & &  \qquad\qquad\qquad\qquad\qquad\qquad
  ~  +  ~
      % \gamma^\zeta ~ S(p -w) ~ 
      \gamma^\mu ~ S(p+k-w) ~\gamma^\nu \bigg\}  \times
      \nonumber \\
      & & \hspace{3cm}  \times  ~ S(p+k+q-w) ~\gamma^{\zeta^\prime} 
%      \nonumber \\
%  & &  \qquad\qquad\qquad
%   + ~
%      \gamma^\zeta ~ S(p -w) ~ \gamma^{\nu}  ~S(p+q-w) ~\gamma^{\zeta^\prime}  ~S(p+q) ~\gamma^\mu \nonumber \\
%  & &  \qquad\qquad\qquad
%   + ~
%      \gamma^\zeta ~ S(p -w) ~ \gamma^{\mu}  ~S(p+k-w) ~\gamma^{\zeta^\prime}  ~S(p+k) ~\gamma^\nu ~
%\bigg\}
    \label{Eq:Gamma-2-2-pert-lowest}
\end{eqnarray}
that represents the box diagrams that are the lowest order in $g$ perturbative contribution to $\Gamma^{\mu\nu}$ and, therefore,
the approximate Dyson-Schwinger equation reproduces the lowest order results of perturbation theory for the two-photon-two-fermion irreducible vertex.
Then, considering the definitions in Eqs (\ref{Eq:GammaPert}) and (\ref{Eq:DSE1-MOM-pert}) and the lowest order
(\ref{Eq:Gamma-2-2-pert-lowest}) for the two-photon-two-fermion irreducible vertex one can use these results in Eq. 
(\ref{Eq:DSE_2photon2fermion-truncated}) to generate the $g^4$ correction to the one-particle irreducible vertex under investigation.
By iterating this procedure further higher order corrections can be computed.
This interactive  procedure requires that renormalization should be performed in an order-by-order approach. 
At the lowest order in the coupling constant, a naive power count shows that the two-photon-two-fermion irreducible vertex is UV and, for massive fermions, 
IR finite. The renormalization of the DSE is discussed in Sec. \ref{Sec:Renor}.

%======================================================
%======================================================
\subsection{Infrared divergences and linear covariant gauges}

A closer look at the loop integrals shows that, in general, $\Gamma^{\mu\nu}$ is combined with a photon propagator as follows
\begin{eqnarray}
& & 
   D_{\zeta\zeta^\prime}(w) \, \Gamma^{\mu\zeta^\prime} (p-w, \, -p-k; \, k, \, w) = \nonumber \\
   & & \hspace{1cm} = ~
   - \, \left( P^\perp \right)_{\zeta\zeta^\prime} (w) \, D(w^2)  \, \Gamma^{\mu\zeta^\prime} (p-w, \, -p-k; \, k, \, w) 
   \nonumber \\
   & & \hspace{2.5cm}
   - \, \frac{\xi}{w^2} \, \left( P^L \right)_{\zeta\zeta^\prime} (w) \, \Gamma^{\mu\zeta^\prime} (p-w, \, -p-k; \, k, \, w) 
\end{eqnarray}
where
\begin{equation}
\left( P^\perp \right)_{\zeta\zeta^\prime} (w) = g_{\zeta\zeta^\prime} - \frac{w_\zeta w_\zeta^\prime}{w^2}
\qquad\mbox{ and }\qquad
\left( P^L \right)_{\zeta\zeta^\prime} (w) = \frac{w_\zeta w_\zeta^\prime}{w^2}
\end{equation}
are the transverse and longitudinal momentum projection operators, respectively. The exception to this rule is the last term appearing
in Eq. (\ref{Eq:DSE_2photon2fermion-truncated}) where $\Gamma^{\mu\zeta^\prime}$ should be replaced by  $\Gamma^{\zeta^\prime}$.
The gauge-dependent part can be simplified with the help of
the Ward-Takahashi identity (\ref{Ward:2photon2fermion}), and for the last term one should use instead Eq. (\ref{Ward:photon-fermion-vertex}).
The contribution of the two-photon-two-fermion to the loop integrals can be replaced by the terms that are proportional to the 
photon-fermion vertex as follows
\begin{eqnarray}
& & 
   D_{\zeta\zeta^\prime}(w) \, \Gamma^{\mu\zeta^\prime} (p-w, \, -p-k; \, k, \, w) = \nonumber \\
   & & \hspace{0.7cm} = ~
   - \, \left( P^\perp \right)_{\zeta\zeta^\prime} (w) \, D(w^2)  \, \Gamma^{\mu\zeta^\prime} (p-w, \, -p-k; \, k, \, w)  \nonumber \\
   & & \hspace{1.2cm}
   - \, \frac{\xi}{w^4} \, w_\zeta \, \bigg( \Gamma^{\mu} (p-w, \, -p-k+w; \, k) - \Gamma^{\mu} (p, \, -p-k; \, k)   \bigg) \ .
\end{eqnarray}
This expression suggests that the problem of the infrared divergences in the linear covariant gauges becomes more difficult when $\xi$ moves
away from zero, i.e. when one departure from the Landau gauge. However, the gauge-dependent part of the last equation can also be viewed
as constraining the vertex for small $w$ where one can do the approximation
\begin{eqnarray}
 & &
- \, \frac{\xi}{w^4} \, w_\zeta \, \bigg( \Gamma^{\mu} (p-w, \, -p-k+w; \, k) - \Gamma^{\mu} (p, \, -p-k; \, k)   \bigg)  \approx  \nonumber \\
& & \hspace{1.6cm}
 \approx 
\frac{\xi}{w^4} \, w_\zeta w_\iota \, \left( \frac{ \partial \Gamma^{\mu} (p, \, -p-k; \, k) }{\partial {p_{1 \, \iota} }  } - \frac{ \partial \Gamma^{\mu} (p, \, -p-k; \, k) }{\partial {p_{2 \, \iota} }  } \right) ,
\label{EQ-IR-1}
\end{eqnarray}
where $\partial \Gamma^\mu / \partial p_j$ refers to differentiation with respect to argument $j$, up to higher order terms in $w$. A naive power count
in $w$ shows that the gauge dependent term is $1/w^2$, reproducing the perturbative like behaviour of the photon
propagator. This power like behavior seems to solve possible infrared divergences that are associated with the gauge\--dependent terms
in the loop integrals. For the type of contributions under discussion, the problem of the infrared divergences in the loop integrals for the gauge
dependent part is solved either if for small momenta the photon-fermion vertex does not depend on the fermion momenta, as is sometimes assumed,
or if the derivative dependence on  both fermion momenta is the same. In both cases the r.h.s. of Eq. (\ref{EQ-IR-1}) vanishes.

Repeating the same argument for the last term in the loop integral of Eq. (\ref{Eq:DSE_2photon2fermion-truncated}), then
\begin{eqnarray}
& & 
- \, \frac{\xi}{w^4} \, w_\zeta \, w_{\zeta^\prime} \, \Gamma^{\zeta^\prime} (p+k+q-w, \, -p-k-q; \, w)  =  
\nonumber \\
& & \hspace{0.8cm} = 
- \, \frac{\xi}{w^4} \, w_\zeta \, \bigg( S^{-1} ( p+k+q)  - S^{-1} (p+k+q-w)  \bigg) \nonumber \\
&  & \hspace{0.8cm} \approx   \frac{\xi}{w^4} \, w_\zeta \, w_\iota \, \frac{\partial S^{-1}(p+k+q)}{\partial p_\iota}  + \cdots
\end{eqnarray}
and the infrared behaviour of the gauge-dependent part reproduces the same type of behaviour as before. 
Inspired on perturbation theory, 
one can  conclude that the infrared divergences for any linear covariant gauges should be of the same nature as the 
possible infrared
divergences that appear in the Landau gauge defined by $\xi = 0$.

We conclude this section noting that
the combination of Eq. (\ref{Eq:DSE_2photon2fermion-truncated}), after contraction with $k_\mu \, q_{\nu}$, with the ``scalar''
Ward-Takahashi  given in Eq. (\ref{Ward:2photon2fermionScalar}) gives a lengthy sum rule that combines the fermion-photon vertex
and the fermion propagator.

%======================================================
%======================================================
\subsection{On the two-photon-two-fermion tensor basis \label{Sec:2P2F-Tensor-Basis}}

The discussion of the solution for the two-photon-two-fermion Ward-Takahashi identity in Sec. \ref{Sec:SolWTI} suggested
a minimal basis of operators to describe, in momentum space, $\Gamma^{\mu\nu}_L$. The minimal basis involves eight different
tensors, see Eq. (\ref{Eq:MinimalBasis-2P2F}), and was built taking into consideration the solution of the WTI and including the set of
linear operators, in the photon momenta, that are associated with the transverse components of the photon-fermion vertex appearing
in the two-photon-two-fermion WTI. However, as discussed in \cite{Tarrach:1975tu,Perrottet:1973qw}, see also \cite{Scherer:1996ux,Eichmann:2012mp}
for particular cases, the most general basis of operators to describe $\Gamma^{\mu\nu}$ requires a larger basis.
For example, in \cite{Tarrach:1975tu} a minimal basis of gauge-invariant tensors with no kinematical singularities was constructed explicitly and
it uses eighteen different operators, that require eighteen different Lorentz scalar form factors. A comparison of
Eq. (\ref{Eq:MinimalBasis-2P2F}) with this explicit construction shows that all the operators reported in Eq. (\ref{Eq:MinimalBasis-2P2F})
also appear in \cite{Tarrach:1975tu}, either explicitly or as linear combinations of what the author calls as $\tau_i$. The minimal basis reported
earlier can be extended to include some transverse components following the procedure devised in \cite{Bardeen:1968ebo} and worked out in
\cite{Tarrach:1975tu} that requires the replacement
\begin{equation}
   \mathcal{O}^{\mu\nu} (k,q) ~ \longmapsto ~ \left( g^{\mu\mu^\prime} - \frac{k^\mu k^{\mu^\prime}}{k^2} \right) 
    \left( g^{\nu\nu^\prime} - \frac{q^\nu q^{\nu^\prime}}{q^2} \right)  ~ \mathcal{O}_{\mu^\prime\nu^\prime} (k,q) \ ,
\end{equation}
where $\mathcal{O}^{\mu\nu} (k,q)$ stands for a generic operator that complies with Bose symmetry, and, when necessary, multiplied by Lorentz invariant
kinematical factors to ensure that the tensor basis is free of kinematical singularities. Even after performing such a construction, the total number of
operators is under eighteen.

%================================================================
%================================================================
\subsection{An approximate solution of the DSE for heavy fermions \label{SecDSE-2P2F-Sol}}

In this section, we would like to discuss an approximate solution to Eq. (\ref{Eq:DSE_2photon2fermion-truncated}). In order to do so, we try to
identify the dominant form factors for $\Gamma_{\mu\nu}$. This is difficult to achieve in a completely satisfactory way, but we turn
our attention to the low momentum solutions given in Eqs (\ref{Eq:WTI-Sol-SoftPhoton-Approx}) and (\ref{Eq:WTI-Sol-Linear}). With the aim
of simplifying the DSE (\ref{Eq:DSE_2photon2fermion-truncated}) we take the heavy fermion limit and consider that
\begin{equation}
  S(p) \approx \frac{1}{m} \ .
\end{equation}  
Then, in the heavy fermion limit the DSE becomes
\begin{eqnarray}
& &
\Gamma^{\mu\nu} (p, \, -p-k-q; \, k, \, q)  ~ \approx \nonumber \\
& &  \quad \approx
   i \, \frac{g^2}{m} ~\int \frac{d^4 w}{(2 \, \pi)^4} ~ D_{\zeta\zeta^\prime}(w) ~\gamma^\zeta   \nonumber \\
        & & \hspace{0.7cm} \bigg\{ 
                     ~ \frac{1}{m^2} ~\Gamma^\mu (p-w, \, -p-k+w; \, k)  ~ 
                     \nonumber \\
                     & & 
                     \hspace{3cm} \Gamma^{\nu} (p+k-w, \, -p-k-q+w; \, q)  
                     \nonumber \\
                     & & \hspace{4cm} 
                                     ~ \Gamma^{\zeta^\prime} (p+k+q-w, \, -p-k-q; \, w)        \nonumber \\
        & & \hspace{0.7cm} ~ + ~ \frac{1}{m^2} ~
                     ~\Gamma^\nu (p-w, \, -p-q+w; \, q)  ~ 
                     \nonumber \\
                     & & \hspace{3cm} \Gamma^{\mu} (p+q-w, \, -p-k-q+w; \, k)  
                     \nonumber \\
                     & &
                     \hspace{4cm}  ~ \Gamma^{\zeta^\prime} (p+k+q-w, \, -p-k-q; \, w)        \nonumber \\
        & & \hspace{0.7cm} ~ + ~ \frac{1}{m} ~
                     ~\Gamma^\mu (p-w, \, -p-k+w; \, k)  ~  
                     \nonumber \\
                     & & \hspace{3cm} ~\Gamma^{\nu\zeta^\prime} (p+k-w, \, -p-k-q; \, q, \, w)        \nonumber \\                     
        & & \hspace{0.7cm} ~ + ~\frac{1}{m} ~
                     ~\Gamma^\nu (p-w, \, -p-q+w; \, q)  ~  
                     \nonumber \\
                     & & \hspace{3cm} ~\Gamma^{\mu\zeta^\prime} (p+q-w, \, -p-k-q; \, k, \, w)       \nonumber \\
        & & \hspace{0.7cm} ~ + ~\frac{1}{m} ~
                     ~\Gamma^{\mu\nu} (p-w, \, -p-k-q+w; \, k, \, q)  ~  
                     \nonumber \\
                     & & \hspace{3cm} ~\Gamma^{\zeta^\prime} (p+k+q-w, \, -p-k-q; \, w)     %  \nonumber \\
%                   & &  XXXXX
~ \bigg\}   \nonumber \\
& &  \quad \approx
   i \, \frac{g^2}{m^2} ~\int \frac{d^4 w}{(2 \, \pi)^4} ~ D_{\zeta\zeta^\prime}(w) ~\gamma^\zeta  ~ \bigg\{ 
   \nonumber \\
   & & \hspace{1.6cm}
                     ~\gamma^\mu   ~  \Gamma^{\nu\zeta^\prime} (p+k-w, \, -p-k-q; \, q, \, w)     
                        \nonumber \\                     
        & & \hspace{2.6cm} 
        ~ + ~
                     \gamma^\nu  ~  \Gamma^{\mu\zeta^\prime} (p+q-w, \, -p-k-q; \, k, \, w)       \nonumber \\
        & & \hspace{3.6cm} ~ + ~ 
                     ~\Gamma^{\mu\nu} (p-w, \, -p-k-q+w; \, k, \, q)  ~  \gamma^{\zeta^\prime}    \bigg\}
  \label{Eq:DSE_2photon2fermion-truncated-K}
\end{eqnarray}
where in the last line we took only the leading term in $1/m$ and set the photon-fermion vertex to its tree level expression.
For the approximation considered here, the terms proportional to the derivatives of $B$ in Eq. (\ref{Eq:WTI-Sol-Linear})
can be ignored and, in this case, the solution of the WTI requires a single tensor operator
\begin{equation}
\Gamma^{\mu\nu}_L (p, \, -p-k-q; \, k, \, q) ~ ~ = ~ ~ - \, \, 2 \, \, A^\prime(p^2) \Big( 
 \slashed{p} \, g^{\mu\nu}  +  p^\mu \, \gamma^\nu \, + \, p^\nu \,\gamma^\mu  \Big)
 \label{Eq:AnsatzSimple}
\end{equation}
and suggests to write in the r.h.s. of the DSE
\begin{eqnarray}
& &
\hspace{-0.4cm}
\Gamma^{\mu\nu} (p, \, -p-k-q; \, k, \, q)  = 
  F_0 \, \Bigg(  \slashed{p} \, g^{\mu\nu}  +  p^\mu \, \gamma^\nu \, + \, p^\nu \,\gamma^\mu  \Bigg) \nonumber \\
  & &  + ~
  F_1 \, \left( g^{\mu\mu^\prime} - \frac{k^\mu k^{\mu^\prime}}{k^2} \right) 
            \left( g^{\nu\nu^\prime} - \frac{q^\nu q^{\nu^\prime}}{q^2} \right) \, 
            \Bigg(  \slashed{p} \, g_{\mu^\prime\nu^\prime}  +  p_{\mu^\prime} \, \gamma_{\nu^\prime} \, + \, p_{\nu^\prime} \,\gamma_{\mu^\prime}  \Bigg) 
            \label{Eq:GammaApp_Simples}
\end{eqnarray}
where $F_0$ and $F_1$ are Lorentz scalar functions of $p^2$, $(p+k+q)^2$, $k^2$ and $q^2$. 
To proceed, let us ignore in Eq. (\ref{Eq:GammaApp_Simples}) the transverse part of $\Gamma^{\mu\nu}$, set 
$F_0 = - 2 A^\prime(p^2)$,  and take the limit $k = q=0$ of
Eq. (\ref{Eq:DSE_2photon2fermion-truncated-K}). Then, this last equation can written as
\begin{eqnarray}
& &
\Gamma^{\mu\nu} (p, \, -p; \, 0, \, 0)  ~ \approx  \,  2 \, i \, \frac{g^2}{m^2} ~\int \frac{d^4 w}{(2 \, \pi)^4} ~ D(w^2) ~  A^\prime( (p-w)^2) \nonumber \\
& & \bigg\{ 
 3 \, \, {g}^{\mu  \nu } \, \, \slashed{p}
~ + ~  {g}^{\mu  \nu } \, \, \slashed{w}
~ -  ~  2   \,\, {\gamma }^{\nu } \, {p}^{\mu }
~ - ~ 2   \, \, {\gamma }^{\mu } \, {p}^{\nu }
~ + ~   {\gamma }^{\mu } \, {w}^{\nu }
 ~ + ~   {\gamma }^{\nu } \, {w}^{\mu } 
\nonumber \\
& & \qquad
 - 4 ~    \frac{ ({p} {w})  }{\, {w}^2} \, \, \slashed{w} \, \, {g}^{\mu  \nu }
 ~ - ~ 5  \,\, \slashed{w} \,  \, \frac{ {p}^{\mu } \, {w}^{\nu }  }{ {w}^2}
 ~ -  ~ 5  \,\,  \slashed{w} \,\, \frac{ {p}^{\nu } \, {w}^{\mu }  }{{w}^2}
  \nonumber \\
  & & \quad
 ~ - ~ 2    \,\, \slashed{p} \,  \,  \frac{ {w}^{\mu } \, {w}^{\nu }}{ {w}^2}
 ~ + ~ 12  \, \, \slashed{w} \,  \, \frac{{w}^{\mu } \, {w}^{\nu } }{ {w}^2}
~ ~  + ~   \frac{({p} {w})}{\, {w}^2}\,\,    {\gamma }^{\mu } \, {w}^{\nu } 
~ + ~    \frac{({p}  {w})}{ {w}^2} \,\,  {\gamma }^{\nu } \, {w}^{\mu } 
\nonumber \\
& & \quad
~ - ~  i \,\, \frac{  \, {w}^{\nu } \, }{\, {w}^2} \, \, {\epsilon }^{\eta  \mu  \alpha\beta} \, {p}_\alpha \, {w}_\beta  \,\, {\gamma }_{\eta }\, {\gamma }_5
~ - ~  i \,\, \frac{ \,  {w}^{\mu } \, }{\, {w}^2} \,\, {\epsilon }^{\eta  \nu  \alpha\beta} \, {p}_\alpha \, {w}_\beta \,\, {\gamma }_{\eta }\, {\gamma }_5 
  \bigg\}
  \label{Eq:DSE_2photon2fermion-truncated-K1}
\end{eqnarray}
and the computation of the two-photon-two-fermion irreducible Green function requires the knowledge of the fermion propagator, i.e. of $A$, that calls
for a solution of the coupled equations considered in the present work or a model of the fermion propagator. However, the analysis of Eq.
(\ref{Eq:DSE_2photon2fermion-truncated-K1}) shows that, in the approximation considered, the tensor structure of the Green function is a linear combination
of the operators
\begin{equation}
  \slashed{p} \,\, g^{\mu\nu}, \qquad
  p^\mu \gamma^\nu + p^\nu \gamma^\mu,  \qquad
  \slashed{p} \,\, p^\mu p^\nu
  \qquad\mbox{ and }\qquad
  i \,\, \epsilon^{\eta\mu\nu\alpha} \,\, p_\alpha \,\, \gamma_\eta \, \gamma_5
\end{equation}
with different coefficients that are functions of $p^2$. The description of $\Gamma^{\mu\nu}$ is certainly more elaborated than the ansatz
(\ref{Eq:AnsatzSimple}) used to feed the r.h.s. of the truncated DSE. 

For the Schwinger model, the authors of \cite{Radozycki:1998pf} also derived a Dyson-Schwinger equation for the two-photon-two-fermion
one particle irreducible Green function and built a self-con\-sis\-ten\-ce solution of the corresponding equation in coordinate space.
We postpone the solution of the above approximate Dyson-Schwinger equation and, therefore, we are unable to comment on similarities and
differences.

%================================================================
%================================================================
\section{Two-photon-two-fermion vertex and its low energy contribution to the effective photon-fermion vertex  \label{Sec:EffVert}}

The Dyson-Schwinger equation for the photon-fermion vertex, see Eq. (\ref{Eq:DSE1-MOM}), shows that this vertex gets a contribution
coming from the two-photon-two-fermion irreducible diagram that reads
\begin{eqnarray}
& & 
   \widetilde{\widetilde{\Gamma}} \, ^\mu (p, \, -p -k; \, k)   =   
   \nonumber \\
   & & =  i \, g^2 \, \int \frac{d^4q}{(2 \, \pi)^4} ~D_{\zeta\zeta^\prime}(q) ~ \Bigg\{
                                \gamma^\zeta \, S(p-q) \, {\Gamma}^{\zeta^\prime\mu} (p - q, \, -p-k; \, q , \, k) 
                                \Bigg\}  \nonumber \\
&  & \approx 
 ~   i \, g^2 \, \int \frac{d^4q}{(2 \, \pi)^4} ~D_{\zeta\zeta^\prime}(q) ~ \Bigg\{
                                \gamma^\zeta \, S(p-q) \, {\Gamma}^{\zeta^\prime\mu} (p - q, \, -p; \, q , \,  0) 
                                \Bigg\}       
                                \nonumber \\
                                & & \qquad         +  ~ \mathcal{O}(k) \ .
                                \label{Eq:EffVertex-2f2f}
\end{eqnarray}
Its low energy limit, defined as the soft photon limit given by taking $k = 0$, 
requires the knowledge of the two-photon-two-fermion vertex in the soft photon limit, whose longitudinal
part can be found in Eq. (\ref{Eq:WTI-Sol-SoftPhoton}). For the Landau gauge, where the photon propagator is transverse, the only terms in
 Eq. (\ref{Eq:WTI-Sol-SoftPhoton}) that contribute to the vertex are
\begin{eqnarray}
& & 
\hspace{-0.4cm}
\Gamma^{\mu\nu} (p, \, -p-k; \, k, \, 0)  = 
\nonumber \\
& & =  2 ~ 
g^{\mu\nu} ~ \Bigg\{  \, \lambda_3 \left( p^2, \, p^2, \, 0 \right)    
~  - ~ 2 \, \slashed{q} \,  \lambda_2 \left( p^2, \, p^2, \, 0 \right)    ~ - ~ 2 \, \slashed{p} \, \lambda_2 \left( p^2, \, p^2, \, 0 \right)    \bigg\}  \nonumber \\
&  & =  2 ~ 
g^{\mu\nu} ~ \Bigg\{  \, B^\prime  \left( p^2 \right)    
~  - ~  \, \slashed{q} \,  A^\prime  \left( p^2  \right)    ~ - ~  \, \slashed{p} \, A^\prime \left( p^2 \right)    \bigg\}
           \label{Eq:WTI-Sol-SoftPhotonXXXX}
\end{eqnarray}
where to write the last line we used Eqs (\ref{EQ:L4}), $A^\prime (x) = d A(x)/dx$, $B^\prime (x) = d B(x)/dx$ and assumed that
the fermion functions $A$ and $B$ are differentiable. Then, inserting the photon propagator decomposition, see Eq. (\ref{PhotonPropagator}),
in Eq. (\ref{Eq:EffVertex-2f2f}) it follows that
\begin{eqnarray}
& & 
   \widetilde{\widetilde{\Gamma}} \, ^\mu (p, \, -p -k; \, k)   =      - \, 2 \, i \, g^2 \, \int \frac{d^4q}{(2 \, \pi)^4} ~ D(q^2) 
        \left( g_{\zeta}^\mu - \frac{q_\zeta q^\mu}{q^2} \right)  ~ 
        \nonumber \\
        & & \qquad\qquad\qquad
                                \gamma^\zeta \, S(p-q) \,    \bigg( B^\prime  \left( p^2 \right)    
  -   \, \left( \slashed{q} +  \slashed{p} \right) \, A^\prime \left( p^2 \right)  \bigg)   + \mathcal{O}(k) \ .
                                \label{Eq:EffVertex-2f2fFFF}
\end{eqnarray}
and the contribution of the two-photon-two-fermion vertex to the photon-fermion vertex can be expressed in terms of the fermion propagator functions.
Significant contributions to the photon-fermion vertex can occur when the derivatives of the $A$ and $B$ become large.
Moreover, the decomposition of Eq. (\ref{Eq:EffVertex-2f2fFFF}) into the Dirac basis shows that the two-photon-two-fermion 
contribution to the photon-fermion vertex results in electric and magnetic dipole-like operators that are proportional to the derivatives of
the fermion propagator functions. Indeed, performing the Dirac algebra, in the limit of $k \rightarrow 0$, 
Eq. (\ref{Eq:EffVertex-2f2fFFF}) results in terms involving  the operators
\begin{equation}
  \gamma^\mu, \qquad   p^\mu,  \qquad \sigma^{\mu\alpha}p_\alpha \quad\mbox{ and }\quad p^\mu \, \slashed{p} \ .
\end{equation}
Corrections in $k$ allow for new classes of operators that are not listed and that require either $A^\prime$, $B^\prime$ or either order
derivatives of these functions.

The analysis of the first term in Eq. (\ref{Eq:DSE1-MOM}) shows that the photon-fermion vertex in soft photon limit
is not determined only by the fermion propagator functions $A$ and $B$ and its derivatives. 
Indeed, the first line in Eq. (\ref{Eq:DSE1-MOM}) includes a contribution of the vertex himself that requires only its soft photon limit and a 
second term that calls for the full vertex. It is only the contribution of the two-photon-two-fermion irreducible diagram that, in the soft photon limit,
can be written in terms of the fermion propagator functions.

%================================================================
%================================================================
\section{Renormalization \label{Sec:Renor}}

The equations derived so far involve only bare quantities and, therefore, it remains to write all the Eqs in term of physical quantities.
In order to do so, one has to go through the renormalization procedure. For completeness, this is worked out in this section.
Herein,  the use of the index ``\textit{phys}''  refers to physical quantities, while the bare quantities have no index. 

For QED the renormalization constants, generically named $Z$, are
\begin{eqnarray}
& &
   A_\mu  = Z^{\frac{1}{2}}_3 \, A^{(phys)}_\mu   , \quad
   \psi  = Z^{\frac{1}{2}}_2 \, \psi^{(phys)} , \quad
   g = \frac{Z_1}{Z_2 \, Z^{\frac{1}{2}}_3}      g^{(phys)}    ,  
   \nonumber \\
   & & 
   m = \frac{Z_0}{Z_2} \, m^{(phys)} \quad\mbox{and}\quad
  \xi = Z_3 \,  \xi^{(phys)}  .
\end{eqnarray}
These definitions imply in the following relations between bare and renormalized propagators 
\begin{equation}
   D_{\mu\nu}(k^2) = Z_3 \,  D^{(phys)} _{\mu\nu}(k^2) 
   \qquad\mbox{ and }\qquad
   S(p) = Z_2 \,  S^{(phys)}(p) 
\end{equation}
and, from the definitions Eqs (\ref{Eq:Vertex_app}) and (\ref{V_2bosons_2fermions}), one has
\begin{equation}
     \Gamma_\mu =  \,  \frac{\Gamma^{(phys)}_\mu}{Z_1}
     \qquad\mbox{ and }\qquad
     \Gamma_{\mu\nu} =  \, \frac{Z_2}{Z^2_1} \,  \Gamma^{(phys)}_{\mu\nu}     \ .
\end{equation}
On the other hand, the Ward identities (\ref{Ward:photon-fermion-vertex})  and (\ref{Ward:2photon2fermion}) imply that
\begin{equation}
   \Gamma_{\mu} = \frac{\Gamma^{(phys)} _{\mu}}{Z_2}
   \qquad\mbox{ and }\qquad
   \Gamma_{\mu\nu} = \frac{\Gamma^{(phys)}_{\mu\nu}}{Z_2} 
\end{equation}
and, therefore, 
\begin{equation}
   Z_1 = Z_2 \ .
\end{equation}
The renormalization program for QED requires three renormalization constants, 
from three renormalization conditions computed from the fermion gap equation (\ref{DSE-Fermion-momentum}), that provide $Z_1 (=  Z_2)$ and $Z_0$,
and the equivalent photon gap equation (\ref{Eq:PhotonEq_LCA}), which determines $Z_3$.
Note that with the above definitions, the coupling constant is renormalized by $\sqrt{Z_3}$ and the combination $g^2 \, D(k^2)$ is independent of the
renormalization scale.

Before proceeding any further, let us write the Dyson-Schwinger equations in their renormalized form where all quantities are finite. To simplify the notation, 
from now on,
in the renormalized equations we will omit the suffix $(phys)$. The renormalized fermion gap equation (\ref{DSE-Fermion-momentum}) reads
\begin{eqnarray}
  S^{-1} (p)  & = &  Z_2 \,   \slashed{p} - Z_0 \, m 
  \nonumber \\
  & & \hspace{0.6cm}
     - \, i \, g^2  \, Z_2 \, \int \frac{d^4 k}{( 2 \, \pi )^4} ~   D_{\mu\nu} (k) ~  \Big[ \gamma^\mu ~ S(p - k) ~   {\Gamma}^\nu (p-k, - p; k)  \Big]    \nonumber  \\
     & = & 
     Z_2 \,   \slashed{p} - Z_0 \, m 
     - \, i \,   g^2 \, \Sigma(p) \ ,
     \label{DSER-gap}
\end{eqnarray}
while the renormalized photon gap equation (\ref{Eq:PhotonEq_LCA}) is given by
\begin{eqnarray}
  \frac{1}{D(k^2)}  & = &
  Z_3 \,  k^2 
  - i \, \frac{g^2}{3} \,  Z_2 \,  \int \frac{d^4 p}{(2 \, \pi)^4} ~ \text{Tr} \Big[ \gamma_\mu \, S(p) \, {\Gamma}^\mu(p, -p + k; -k ) \, S(p - k ) \Big] \nonumber \\
  & = & Z_3 \,  k^2 
  - i \,  \frac{g^2}{3} \,   k^2 \, \Pi(k^2) 
     \label{DSER-photon}
\end{eqnarray}
where $\Sigma (p)$ and $\Pi (k^2)$ are the fermion and the photon self-energies, respectively.
Then, the renormalized Dyson-Schwinger equation for the vertex (\ref{Eq:DSE1-MOM}) is
\begin{eqnarray}
& & 
   {\Gamma}^\mu (p, \, -p -k; \, k)   =  Z_2 \, \gamma^\mu  ~ + ~ i \, g^2 \,  Z_2 \, \int \frac{d^4q}{(2 \, \pi)^4} ~D_{\zeta\zeta^\prime}(q) \nonumber \\
   & & \Bigg\{
                                \gamma^\zeta \, S(p-q) \, {\Gamma}^\mu ( p - q, \, -p -k + q; \, k) \,
                                S(p+k-q) \, {\Gamma}^{\zeta^\prime} (p +k-q, \, -p-k; q) 
                                 \nonumber \\
   & & \qquad \qquad~ + ~
                                \gamma^\zeta \, S(p-q) \, {\Gamma}^{\zeta^\prime\mu} (p - q, \, -p-k; \, q , \, k) 
                                \Bigg\}  \nonumber \\
  &  & =  ~      Z_2 \, \gamma^\mu  ~ + ~ i \, g^2 \,  \Delta \Gamma^\mu (p,k)    \ .
                                \label{DSER-vertex1}
\end{eqnarray}

It is convenient to introduce some extra notation and write the inverse of the renormalized fermion propagator as
\begin{equation}
   S^{-1}(p) = A(p^2) \, \slashed{p} - B(p^2)
   \label{Eq:FPropInv_A_B}
\end{equation}   
and, therefore,
\begin{equation}
   S(p) = Z(p^2) \, \frac{\slashed{p} + M(p^2)}{p^2 - M^2(p^2) + i \, \epsilon} 
   \label{Eq:FProp_A_B}
\end{equation}   
with $Z(p^2) = 1 / A(p^2)$ and $M(p^2) = B(p^2) / A(p^2)$. It is also common practice to write the fermion self-energy as
\begin{equation}
   \Sigma (p) = \Sigma_v(p^2) \, \slashed{p} + \Sigma_s(p^2) \ ,
\end{equation}
that allows to split the fermion gap equation into a vectorial and a scalar part as follows
\begin{eqnarray}
 & &
  A (p^2)  = Z_2  \,  
     - \, i \,  \frac{g^2}{p^2} \,  Z_2 \,  \int \frac{d^4 k}{( 2 \, \pi )^4} ~   D_{\mu\nu} (k) ~  
                                   \text{Tr}  \Big[  \, \slashed{p} ~ \gamma^\mu ~ S(p - k) ~   {\Gamma}^\nu (p-k, - p; k) \,  \Big] 
                                   \nonumber \\
                                   & & \hspace{1cm}
     =  Z_2  - \, i \,  g^2 \, \Sigma_v(p^2) \ , 
     \label{DSER-gap-vec} \\
     & &
  B (p^2)  =  Z_0 \, m 
     + \, i \,   g^2 \,  Z_2 \, \int \frac{d^4 k}{( 2 \, \pi )^4} ~   D_{\mu\nu} (k) ~   
                                    \text{Tr} \Big[  \, \gamma^\mu ~ S(p - k) ~   {\Gamma}^\nu (p-k, - p; k) \,  \Big] 
                                    \nonumber \\
                                    & & \hspace{1cm}
     = Z_0 \, m + i \,   g^2 \, \Sigma_s(p^2)  \ .
     \label{DSER-gap-sca}
\end{eqnarray}

One way to define the renormalization constants is choosing renormalization conditions implemented using
 Eqs (\ref{DSER-photon}), (\ref{DSER-gap-vec}) and (\ref{DSER-gap-sca}). Thus,  demanding that
\begin{equation}
A(\mu^2) = 1, \qquad B(\mu^2) = m \qquad\mbox{ and }\qquad D(\mu^2) = \frac{1}{\mu^2} 
  \label{Renormalization-Conditions}
\end{equation}
it comes that
\begin{eqnarray}
Z_2  & = &  1 \, + \, i \,  g^2 \, \Sigma_v( \mu^2 )   \ , \\
Z_0 & = & 1 \,  -  i \,   g^2 \, \Sigma_s(\mu^2) / m \ ,  \\
Z_3 & = & 1 \, + \, \frac{i}{3} \, g^2 \, \Pi(\mu^2) ,
\end{eqnarray}
recall that $\Sigma_v$, $\Sigma_s$, $\Pi$ include a $Z_2$ factor in its definition, and the renormalized equations for the propagators are
\begin{eqnarray}
 & &
  A (p^2)       =  1  - \, i \,  g^2 \, \bigg( \Sigma_v(p^2) - \Sigma_v(\mu^2) \bigg) \ , 
     \label{DSERF-gap-vec} \\
     & &
  B (p^2)  =   \, m + i \,   g^2 \, \bigg( \Sigma_s(p^2) - \Sigma_s(\mu^2) \bigg) \ , 
     \label{DSERF-gap-sca} \\
     & &
    \frac{1}{D(k^2)}  = k^2 \bigg( 1 - \frac{i}{3} \, g^2 \, \Big( \Pi(k^2) - \Pi (\mu^2) \Big) \bigg) 
     \label{DSERF-photon} \ . 
\end{eqnarray}
Here, we have used the same renormalization scale for the fermion and the photon, but one could use instead different mass scales for the renormalization of
fermionic and bosonic fields.

%================================================================
%================================================================
\section{Summary, Dicussions and Conclusions \label{Sec:Summary}}

In the current work, the Dyson-Schwinger equations in Minkowski spacetime are considered to address QED in a general linear covariant gauge.
The DSE equations are an infinite tower of integral equations and, therefore, only  approximate solutions can be built by dealing 
with truncated versions of the full set of equations. A minimal set of integral equations to tackle both
the propagators and the photon-fermion vertex in QED is provided. 
These equations for the two-point and three-point Green functions are exact but, to build a closed set of equations,
one also consider a truncated Dyson-Schwinger equation for the two-photon-two-fermion one-particle irreducible vertex.

The Dyson-Schwinger equations can, in principle, be solved for any regime of the theory that includes the small and strong
coupling regimes and, therefore, allow to solve QED beyond perturbation theory. 
The analysis of the truncated two-photon-two-fermion DSE in a perturbative-type of solution, 
shows that it reproduces the lowest order perturbative solution, as expected.
Moreover, by combining the various integral equations derived, it is possible to build an iterative procedure that allows to compute
higher order corrections in the coupling constant, mimicking the perturbative expansion.
More, we were able to solve the truncated integral equation for $\Gamma^{\mu\nu}$ for a particular kinematical limit, in the heavy quark model,
feeding the equation with a simplified model of this vertex.
The closed set of equations considered allow to investigate the solutions of QED for the propagators, for the photon-fermion vertex
and for the two-photon-two-fermion vertex and to study their gauge dependence. 
In particular, the study of the gauge dependence can be confronted with the results of the Landau-Khalatnikov-Fradkin transformations
\cite{Landau:1955zz,Fradkin:1955jr,DeMeerleer:2018txc,DeMeerleer:2019kmh,DallOlio:2021njq}
as a test to the truncation itself.

Along with the derivation of the Dyson-Schwinger equations, the Ward-Takahashi identities for the photon-fermion and  the
two-photon-two-fermion vertices are also derived. 
As discussed, the WTI for the photon-fermion vertex fixes its longitudinal component relative to the photon momentum, providing 
the so-called Ball-Chiu vertex. 
In the Ball-Chiu vertex the form factors are functions of the fermion propagators.
The compatibility of the DSE for the vertex (\ref{Eq:DSE1-MOM}) with the WTI 
(\ref{Ward:photon-fermion-vertex}) is, in practise, difficult to achieve in a solution of the vertex equation. 
Given that the WTI results from the gauge symmetry directly, set the longitudinal part of the vertex, a solution of (\ref{Eq:DSE1-MOM}) 
that does not spoils gauge symmetry can be built by considering only its transverse component, i.e. by taking (\ref{EQ:L1}) - (\ref{EQ:L4}) 
for $\Gamma^\mu_L$ and solve the equation resulting from contracting (\ref{Eq:DSE1-MOM}) with the transverse operator 
$P^\perp_{\mu\nu}(k)$.
In this way, a dynamical equation for the transverse part of the photon-fermion vertex is engineered keeping its longitudinal part 
compatible with the WTI.

The Ward-Takahashi identity for the two-photon-two-fermion vertex calls for the full photon-fermion vertex and, this vertex being more complex
that the fermion propagators, make its analysis rather involved. 
However, we are able to solve this WTI exactly for the longitudinal component, in the low photon momenta limit.
For the soft photon limit, the solution is given in Eq. (\ref{Eq:WTI-Sol-SoftPhoton}) and writes $\Gamma^{\mu\nu}_L$ in terms of the photon-fermion 
vertex longitudinal form factors, that are functions of the fermion propagator functions $A$ and $B$ themselves. 
A solution of the WTI keeping only its linear terms in the photon momenta can be found in
(\ref{Eq:WTI-Sol-Linear}), suggesting a minimal basis to describe the two-photon-two-fermion vertex.
Further, the WTI for any kinematics is solved and its solution determines $\Gamma^{\mu\nu}_L$ for a general configuration of momenta,
see Eqs (\ref{Sol:WTITwoPhotonTwoFermion1}) and (\ref{Sol:WTITwoPhotonTwoFermion3}).
In what concerns the transverse part of $\Gamma^{\mu\nu}$, the  proposal is to computed it by solving the truncated DSE for the vertex, i.e.
Eq. (\ref{Eq:DSE_2photon2fermion-truncated}) and, indeed, an approximate solution is built that include further operators than the original
ansatz used to build the solution. 

The compatibility of the DSE for the two-photon-two-fermion vertex, i.e. Eq. (\ref{Eq:DSE_2photon2fermion-truncated}),
with gauge symmetry, expressed through the WTI in Eq. (\ref{Ward:2photon2fermion}), is similar to the computation of the solution for
the vertex equation (\ref{Eq:DSE1-MOM}) that is compatible with the WTI (\ref{Ward:photon-fermion-vertex}). Once more this can be
solved by setting $\Gamma^{\mu\nu}_L$ as given by the solution of the two-photon-two-fermioin WTI identity, and replacing Eq. 
(\ref{Eq:DSE_2photon2fermion-truncated}) by its transverse part that results from contractions with projection operators
\begin{equation}
\Gamma_{T}^{\mu\nu} (p, -p-k-q; k, q) = P^\perp \, ^{\mu} \, _{\mu^\prime}(k) ~~  P^\perp \, ^{\nu} \, _{\nu^\prime}(q) ~~
 \Gamma^{\mu^\prime\nu^\prime}(p, -p-k-q; k, q)
\end{equation}

The complexity of the closed set of integral equations prevent us from attempting to solve the set of coupled equations
for the various Green functions. However, a perturbative treatment of the equations that were derived recover, at least to lhe lowest order
in the coupling constant, the results of perturbation theory.
Furthermore, we recall the reader that by solving the Ward-Takahashi identities the  longitudinal parts of the photon-fermion 
and two-photon-two-fermion vertices are determined exactly.
The computation of solutions of the integral equations, together with its phenomenological implications, 
explored via the corresponding Bethe-Salpeter equation, Faddeev equations, etc., will be the subject of future publications, .

%================================================================
%================================================================
\section*{Acknowledgments}

This work was partly supported by the FCT – Funda\c{c}\~ao para a Ci\^encia e a Tecnologia, I.P., under Projects Nos. UIDB/04564/2020 and UIDP/04564/2020.
H. L. Macedo acknowledges financial support via the Research Fellowship - LUGUS 781687 within the R\&D Unit 
CFisUC - Centro de F\'{\i}sica de Universidade de Coimbra, reference: UIDB/04564/2020, financed by 
FCT - Funda\c{c}\~ao para a Ci\^encia e Tecnologia.
R. C. Terin acknowledges the Technological Institute of Aeronautics (ITA) for the hospitality during his Post-Doc financed by the National Council for Scientific and Technological Development (CNPq) being supervised at that period by Prof. Tobias Frederico (ITA).

%================================================================
%================================================================
\appendix

%================================================================
%================================================================
\section{Decomposing the connected Green's functions \label{Sec:Dec}}

In this appendix, we discuss the decomposition of the connected Green functions that are required in the current work. 
Starting from the last relation given in  the set of  Eqs (\ref{DeltaGammas}) and taking its second derivative with respect to $A_{c}$ after $\eta$,
followed by some algebra, and after setting the sources to zero, one arrives at
\begin{eqnarray}
& &
\int d^4 w_1 \, d^4 w_2  ~ \frac{\delta^2 \Gamma}{\delta\psi_{cl, \, \iota}(w_1) \, \delta\bar\psi_{cl, \, \alpha} (x) } \,
         \frac{\delta^2 \Gamma}{\delta A_{cl, \, \nu}(z) \, \delta A_{cl, \, \nu^\prime} (w_2) } \,
          \frac{\delta^3 W}{\delta J^{\nu^\prime}(w_2) \, \delta\eta_{\beta} (y) \, \delta\bar\eta_\iota (w_1) } = \nonumber \\
& &
\hspace{1cm}    
 =       
\int d^4 w  ~ \frac{\delta^3 \Gamma}{ \delta A_{cl, \, \nu}  (z) \, \delta\psi_{cl, \, \iota}(w) \, \delta\bar\psi_{cl, \, \alpha} (x) } \,
          \frac{\delta^2 W}{ \delta\eta_{\beta} (y) \, \delta\bar\eta_\iota (w) } \ .
\end{eqnarray}
This equation can be solved for the connected Green function using the orthogonality relations (\ref{OrthoA}) - (\ref{Ortho2}) and gives
\begin{eqnarray}
& & 
\frac{\delta^3 W}{\delta J_{\mu}(z) ~ \delta\eta_{\beta} (y) ~ \delta\bar\eta_\alpha (x) }   =   \int  d^4w_1 ~ d^4w_2 ~ d^4w_3
  \nonumber \\
& &
 \qquad\qquad\qquad
             \frac{\delta^2 W}{\delta J_{\mu}(z) \, \delta J^{\mu^\prime} (w_1)} \,
            \frac{\delta^2 W}{ \delta\eta_{\alpha^\prime} (w_2) \, \delta\bar\eta_\alpha (x) } 
            \frac{\delta^2 W}{ \delta\eta_{\beta} (y) \, \delta\bar\eta_{\beta^\prime} (w_3) } \,
            \nonumber \\
             & & \qquad\qquad\qquad
            \frac{\delta^3 \Gamma}{ \delta A_{cl, \, \mu^\prime}  (w_1) ~ \delta\psi_{cl, \, \beta^\prime}(w_3) ~ \delta\bar\psi_{cl, \, \alpha^\prime} (w_2) } 
            \label{Decomp3cV0}
\end{eqnarray}
that in terms of the full fermion propagator (\ref{quarkprop}) and of the full gauge boson propagator (\ref{Eq:PhotonProp}), 
i.e. after setting the sources to zero, reads
\begin{eqnarray}
& &
\frac{\delta^3 W}{\delta J^{\mu}(z) \, \delta\eta_{\beta} (y) \, \delta\bar\eta_\alpha (x) }    =   \nonumber \\
& & 
  = ~ g \, \int \, d^4w_1 \, d^4w_2 \, d^4w_3 ~
     D_{\mu\mu^\prime}(z -w_1) ~  \bigg[ S (x - w_2 ) ~  {\Gamma}^{\mu^\prime} (w_2, w_3 ; w_1) ~  S( w_3 - y)  \bigg]_{\alpha\beta}
     \label{Dec3Point}
\end{eqnarray}
where  the photon-fermion vertex was defined as
\begin{equation} 
\left. 
\frac{\delta^3 \Gamma}{ \delta A_{c, \, \mu}  (z) \, \delta\psi_{c, \, \beta} (y) \, \delta\bar\psi_{c, \, \alpha} (x) } 
 \right|_{J=\bar\eta=\eta=0}
= 
 - \, g \, \Big( {\Gamma}^\mu \Big)_{\alpha\beta} (x, y; z) \ .
 \label{Eq:Vertex_app}
\end{equation}
In momentum space, given the definitions in Eqs (\ref{Eq:SFermion_mom}), (\ref{Eq:Dphoton_mom}), and (\ref{Eq:Vert_mom}),
the decomposition of the photon-fermion irreducible vertex is written as
\begin{eqnarray}
\frac{\delta^3 W}{\delta J^{\mu}(z) \, \delta\eta_{\beta} (y) \, \delta\bar\eta_\alpha (x) }    & = & 
       g \, \int \, \frac{d^4 k}{(2 \, \pi)^4} \, \frac{d^4 p^\prime}{(2 \, \pi)^4} \, \frac{d^4 p}{(2 \, \pi)^4} \quad
      e^{- \, i \, k \, z ~ - \, i \,  p^\prime \, x ~ + \, i \, p \, y} \quad
      ( 2 \, \pi )^4 \, \delta(p^\prime - p + k )  ~ \nonumber \\
      & & \hspace{2cm} 
     D_{\mu\mu^\prime}(k) ~ \Big[  S ( p^\prime ) ~  {\Gamma}^{\mu^\prime} ( p^\prime, \, - \, p; \, k ) ~
     S( p)  \Big]_{\alpha\beta} \ .
     \label{Dec3Point-mom}
\end{eqnarray}

The Dyson-Schwinger equation for the photon-fermion vertex (\ref{VDSE}) calls for the  three-point connected Green function, see Eq. (\ref{Dec3Point}),
and the four-point connected Green with two-fermion and two-gauge boson external lines. 
The latter Green function can be decomposed in terms of irreducible functions 
following the same procedure as for the three-point photon-fermion correlation function. 
Starting from Eq. (\ref{Decomp3cV0})  and  performing the
required derivatives, after some straightforward algebra and after setting all the sources to zero,
 one arrives at the decomposition of the four-point correlation function in terms of one-particle irreducible functions 
\begin{eqnarray}
& & 
\frac{\delta^4 W}{\delta J^\mu (z) \, \delta J^\nu (w) \, \delta \eta_\beta (y) \, \delta \bar\eta_\alpha (x) } = \nonumber \\
& &
\qquad  = ~   ~  - ~  g^2 \, 
 \int d^4u_1 \, d^4u_2 \, d^4u_3 \, d^4u_4 ~~ 
                          D_{\mu \mu^\prime} (z - u_1) \, 
                          D_{\nu \nu^\prime} (w - u_2) \nonumber \\
                          & & \hspace{3cm}
                          \Big[ 
                          S(x - u_3) ~  {\Gamma}^{\mu^\prime\nu^\prime}  (u_3, u_4; u_1, u_2) 
                                S(u_4 - y ) \Big]_{\alpha\beta}  \nonumber \\
& & 
\qquad                           ~
 - ~  g^2 \, \int d^4u_1 \, d^4u_2 \, d^4u_3 \, d^4u_4 \, d^4u_5 \, d^4u_6 ~~
           D_{\mu\mu^\prime}(z - u_1) \,
           D_{\nu\nu^\prime}(w - u_2)  ~ D_{\bar\nu\bar\mu}( u_4 - u_3)   \nonumber \\
           & & \hspace{3cm}
           \Big[ S(x - u_5) ~  \Gamma^{\bar\nu} (u_5, u_6; u_4) ~ S( u_6 - y )  \Big]_{\alpha\beta}  ~~
           {\Gamma}^{\mu^\prime\nu^\prime\bar\mu} (u_1, u_2, u_3)
  \nonumber \\
  & &
  \qquad
   - ~ g^2 \,  \int d^4u_1 \, d^4u_2 \, d^4u_3 \, d^4u_4 \, d^4u_5 \, d^4u_6 ~~
                   D_{\mu\mu^\prime}( z - u_1) \, D_{\nu\nu^\prime}(w - u_2)  \nonumber \\
                   & & \hspace{2cm}
                   \Big[ S(x - u_3) ~  {\Gamma}^{\nu^\prime} (u_3, u_5; u_2) ~
                   S(u_5 - u_6) ~ {\Gamma}^{\mu^\prime} (u_6, u_4; u_1) ~
                   S(u_4 - y) \Big]_{\alpha\beta}
  \nonumber \\
  & &
  \qquad
   - ~  g^2 \, \int d^4u_1 \, d^4u_2 \, d^4u_3 \, d^4u_4 \, d^4u_5 \, d^4u_6 ~~
                   D_{\mu\mu^\prime}( z - u_1) \, D_{\nu\nu^\prime}(w - u_2)  \nonumber \\
                   & & \hspace{2cm}
                   \Big[ 
                   S(x - u_3) ~ {\Gamma}^{\mu^\prime} (u_3, u_5; u_1) ~
                   S(u_5 - u_6) ~  {\Gamma}^{\nu^\prime} (u_6, u_4; u_2) ~
                   S (u_4 - y) \Big]_{\alpha\beta}
   \label{Dec4Point}
\end{eqnarray}
where we have used the relation
\begin{eqnarray}
& & 
\frac{\delta^3 W}{\delta J^\mu (x) ~ \delta J^\nu (y) ~\delta J^\zeta (z)}   = 
 g ~ \int d^4 u_1 \, d^4 u_2 \, d^4 u_3 ~ \nonumber \\
 & & \hspace{2cm}
    D_{\mu\mu^\prime} (x - u_1) ~ D_{\nu\nu^\prime} (y - u_2) ~ D_{\zeta\zeta^\prime} (z - u_3) ~
    \Gamma^{\mu^\prime\nu^\prime\zeta^\prime}( u_1, u_2, u_3) 
\end{eqnarray}
that follows from taking functional derivatives of Eq. (\ref{OrthoA}), using this orthogonality relation and setting the sources to zero, and where the notation
\begin{eqnarray}
   \frac{\delta^4\Gamma}{\delta A_{cl, \, \mu} (x) ~ \delta A_{cl, \, \nu} (y) ~ \delta\psi_{cl, \, \beta} (z) ~ \delta\bar\psi_{cl, \, \alpha} (w)} 
        & = &  ~ - \, g^2 \, \Big( {\Gamma}^{\mu\, \nu} \Big)_{\alpha\beta} ( w, z; x, y)  \ ,
        \label{V_2bosons_2fermions} \\
   \frac{\delta^3 \Gamma}{\delta A_{cl, \, \mu} (x) ~ \delta A_{cl, \, \nu} (y) ~ \delta A_{cl, \, \zeta} (z) } 
        & = & ~ - \, g ~ {\Gamma}^{\mu\,\nu\,\zeta}( x, y, z)  
        \label{V_3bosons} 
\end{eqnarray}
was used, together with the bose symmetry of the vertices, i..e, that the vertices with multiple bosonic lines are symmetric under interchange of pair of indices. 
In QED, Furry's theorem tell us that the three photon vertex vanishes and the contribution proportional to 
$\Gamma^{\mu\nu\zeta}$ can be ignored.
Following the same rules as those used to write Eq. (\ref{Dec3Point}) in momentum space, see Eq. (\ref{Dec3Point-mom}),  the decomposition of
the connected four-point Green function (\ref{Dec4Point}) in momentum space reads
\begin{eqnarray}
& & 
\frac{\delta^4 W}{\delta J^\mu (z) \, \delta J^\nu (w) \, \delta \eta_\beta (y) \, \delta \bar\eta_\alpha (x) } ~ =  \nonumber \\
& & 
= ~ -  ~
g^2 \, \int \,  \frac{d^4 k}{(2 \, \pi )^4} \,  \frac{d^4 q}{(2 \, \pi )^4} \,  \frac{d^4 p^\prime}{(2 \, \pi )^4} \,  \frac{d^4 p}{(2 \, \pi )^4} ~
\nonumber \\
& & \qquad
e^{ - \, i \, k \, z ~ - i \, q \, w ~ - \, i \, p^\prime \, x ~ +  \,  i \, p \, y}  ~( 2 \, \pi)^4 ~\delta( p^\prime - p + q + k ) ~~
D_{\mu\mu^\prime}(k) \, D_{\nu\nu^\prime} (q)  \nonumber \\
& &
\qquad
\Bigg\{ ~ \Big[ S(p^\prime) ~ {\Gamma}^{\nu^\prime\mu^\prime}(p^\prime, - p; q, k) ~ S(p) \Big] _{\alpha\beta} \nonumber \\
%& &
%   \hspace{1.5cm} + ~
%    {\Gamma}^{\nu^\prime\mu^\prime\zeta}(q, k, -q-k) ~~ D_{\zeta\zeta^\prime}( p^\prime - p) ~~
%    \Big[ S(p^\prime) ~ {\Gamma}^{\zeta^\prime}(p^\prime, - p; p - p^\prime) ~S(p) \Big] _{\alpha\beta}  \nonumber \\
 & & \hspace{1.3cm} 
    + ~ \Big[ S(p^\prime) ~ {\Gamma}^{\nu^\prime}(p^\prime, - p^\prime - q; q) ~ S(p^\prime + q)  ~ 
                  {\Gamma}^{\mu^\prime}(p^\prime+q, - p ; k) ~S(p) \Big] _{\alpha\beta}  \nonumber \\     
 & & \hspace{1.3cm} 
    + ~ \Big[ S(p^\prime) ~ {\Gamma}^{\mu^\prime}(p^\prime, - p^\prime - k; k) ~ S(p^\prime + k)  ~ 
                  {\Gamma}^{\nu^\prime}(p-q, - p ; q) ~S(p) \Big] _{\alpha\beta}     ~\Bigg\} \ .
    \label{Dec4Point-mom}
\end{eqnarray}
Recall that Eqs (\ref{Dec4Point}) and (\ref{Dec4Point-mom}) are valid only for $J = \eta = \bar\eta = A_c = \psi_c = \bar\psi_c = 0$. 

%================================================================
%================================================================
\section{Ward-Takahashi identities in momentum space \label{Sec:WTI-Mom}}

The Ward-Takahashi identities (\ref{Ward1}) and (\ref{Ward2}) are given in coordinate space, whose translation into momentum space
 (\ref{Ward:photon-fermion-vertex}) and (\ref{Ward:2photon2fermion}) is discussed now.
Starting with the WTI for the photon-fermion vertex (\ref{Ward1}) 
that  in terms of the connected Green functions, i.e. with respect to derivatives of the generating functional $W$, 
see Eq. (\ref{Eq:Geradores}) for definitions, reads
\begin{eqnarray}
& & 
 \frac{1}{g \, \xi} \, \square_x \, \partial^\mu_x \, \left( \frac{\delta^3  W}{ \delta J^\mu(x) \, \delta\eta_\beta (z) \, \delta\bar\eta_\zeta (y) } \right)
 \nonumber \\
 & &  \hspace{2cm} 
 - \, i \, \delta(x- y)   ~ S_{\zeta\beta} ( x - z) 
 + \, i \,  \delta(x- z)   ~ S_{\zeta\beta} ( y - x )  = 0 \ ,
  \label{Ward2Connected}
\end{eqnarray}
where we have used the definition of the fermion propagator (\ref{quarkprop}), and considered the decomposition of the three point function given in Eq.
(\ref{Dec3Point-mom}), the photon propagator in momentum space (\ref{PhotonPropagator}) and the Fourier decomposition of the fermion propagator
(\ref{Eq:SFermion_mom}) one arrives, after some algebra, at
\begin{eqnarray}
& &
\int \frac{d^4k}{(2 \, \pi)^4}\, \frac{d^4p^\prime}{(2 \, \pi)^4} \, \frac{d^4p}{(2 \, \pi)^4} ~ 
    e^{- i \, kx \, -i \, p^\prime y \, -  i  \, p z } ~ (2 \, \pi)^4 \, \delta(p^\prime - p + k ) ~
              k_\mu \Big[ S(p^\prime) \, \Gamma^\mu (p^\prime, \, - p ; \, k ) \, S(p) \bigg]  \nonumber \\
& &
\hspace{1cm} + ~
\int  \frac{d^4p^\prime}{(2 \, \pi)^4} \, \frac{d^4p}{(2 \, \pi)^4} ~e^{- i \, p(x - y)  \, -i \, p^\prime (y  - z)  } ~ S(p^\prime)  ~   \nonumber \\
& &
\hspace{2cm} - ~
\int  \frac{d^4p^\prime}{(2 \, \pi)^4} \, \frac{d^4p}{(2 \, \pi)^4} ~e^{- i \, p(x - z)  \, -i \, p^\prime (y  - x)  } ~ S(p^\prime)  ~   = ~ 0 \ .
\end{eqnarray}
Then, performing an inverse Fourier transformation and multiplying the resulting equation by the inverse of fermion propagator one arrives at Eq. (\ref{Ward:photon-fermion-vertex}). 

The rewriting of the two-photon-two-fermion WTI in momentum space follows  the same steps 
as those performed for the photon-fermion identity.  In terms of the connected Green functions, Eq. (\ref{Ward3}) is given by
\begin{eqnarray}
& & 
 \frac{1}{g \, \xi} \, \square_x \, \partial^\mu_x \, 
           \left( \frac{\delta^4  W}{ \delta J^\nu(w) \, \delta J^\mu(x) \, \delta\eta_\beta (z) \, \delta\bar\eta_\zeta (y) }  ~ + ~ i \, D_{\mu\nu}(x -w) \, S_{\zeta\beta}(y-z)   \right) 
           \nonumber \\
  & & 
   \qquad
            - \frac{i}{g} ~ \partial^\mu_x \Big( g_{\mu\nu} \, \delta (x - w ) \Big) \,  S_{\zeta\beta}(y-z) 
  \nonumber \\
 & & 
 \qquad\qquad
 + ~  i \, \delta(x- y)   \left( \frac{\delta^3 W }{\delta J^\nu (w) \, \delta\eta_\beta (z) \, \delta\bar\eta_\zeta (x) }  \right) ~
 \nonumber \\
  & &
  \qquad\qquad\qquad
 -  ~  i \,  \delta(x- z)   \left( \frac{\delta^3  W }{\delta J^\nu (w) \, \delta\eta_\beta (x) \, \delta\bar\eta_\zeta (y) }  \right) = 0  
  \label{Ward3Connected}
\end{eqnarray}
where we have used the definition of the fermion propagator, see Eq. (\ref{quarkprop}). 
From the Fourier decomposition of the photon propagator (\ref{Eq:Dphoton_mom}) it follows that
\begin{equation}
\square_x \, \partial^\mu_x \, D_{\mu\nu}( x - w) = \xi \, \partial_\mu^x \bigg( \delta (x -w ) \bigg) 
\end{equation}
and, therefore, the second and third terms in Eq. (\ref{Ward3Connected}) cancel exactly. Then, the WTI simplifies into
\begin{eqnarray}
& & 
 \frac{1}{g \, \xi} \, \square_x \, \partial^\mu_x \, 
           \left( \frac{\delta^4  W}{ \delta J^\nu(w) \, \delta J^\mu(x) \, \delta\eta_\beta (z) \, \delta\bar\eta_\zeta (y) }   \right) 
  \nonumber \\
 & & 
 \qquad\qquad
 + ~  i \, \delta(x- y)   \left( \frac{\delta^3 W }{\delta J^\nu (w) \, \delta\eta_\beta (z) \, \delta\bar\eta_\zeta (x) }  \right) ~
 -  ~  i \,  \delta(x- z)   \left( \frac{\delta^3  W }{\delta J^\nu (w) \, \delta\eta_\beta (x) \, \delta\bar\eta_\zeta (y) }  \right) 
 \nonumber \\
 & & \qquad\qquad\qquad\qquad
 = 0  \ .
  \label{Ward3Connected-1}
\end{eqnarray}
Inserting the decompositions of the four point vertex (\ref{Dec4Point-mom}) and of the three point vertex (\ref{Dec3Point-mom}), 
after some straightforward algebra  and after performing an inverse Fourier transformation, this equation reduces to
\begin{eqnarray}
& & 
 D_{\nu\nu^\prime}(q) ~k_\mu ~ S(p) ~ \Big\{
        ~ \Gamma^\mu ( p, \, - p - k; \, k ) ~ S(p+k) ~ \Gamma^{\nu^\prime} (p+k, \, -p-k-q; \, q) \nonumber \\
        & & \hspace{3cm}
        + ~  \Gamma^{\nu^\prime} ( p, \, -p-q; \, q) ~ S(p+q) ~   \Gamma^\mu ( p + q, \, -p-k-q; \, k ) \nonumber \\
        & & \hspace{3cm}
        + ~   \Gamma^{\mu\nu^\prime} (p, \, -p-k-q; \, k, \, q) ~ \Big\}  ~S(p+k+q) \nonumber \\
  & & \qquad
     + ~   D_{\nu\nu^\prime}(q) ~  S(p+k) ~ \Gamma^{\nu^\prime} ( p+k, \, -p-k-q; \, q)  ~ S(p+k+q)  \nonumber \\
  & & \qquad
  -  ~  D_{\nu\nu^\prime}(q) ~  S(p) ~\Gamma^{\nu^\prime} (p, \, -p-q; \, q ) ~ S( p+q)  ~ = ~ 0 \ . 
\end{eqnarray}
When rewriting Eq. (\ref{Ward3Connected}) we considered Furry theorem and
ignored the term proportional to the three-photon irreducible vertex.
Then, multiplying the above expression by the inverse of the photon propagator, the inverse of the fermion propagators, after using the WTI 
(\ref{Ward:photon-fermion-vertex}) to simplify the contractions of $k_\mu$ with the photon-fermion vertex, one arrives at
\begin{equation}
        \Gamma^{\nu} ( p+k, \, -p-k-q; \, q)  ~ - ~ \Gamma^\nu ( p, \, - p - q; \, q ) 
        ~  + ~   k_\mu \, \Gamma^{\mu\nu} (p, \, -p-k-q; \, k, \, q)  ~ = ~ 0 \ . 
\end{equation}
that is the momentum space WTI (\ref{Ward:2photon2fermion} )appearing in the main text.
We call the reader attention that the derivation of these expressions requires explicitly the use of the inverse of the
photon propagator that in the Landau gauge and within this formalism is not defined.

%%%%%%%%%%%%%%%%%%%%%%%%%   Bibliography   %%%%%%%%%%%%

\end{document}